\documentclass[journal]{IEEEtran}
\ifCLASSINFOpdf
   \usepackage[pdftex]{graphicx}
\else
\fi
\usepackage{amsmath,amssymb,amsfonts}
\usepackage{amsthm}

\usepackage[linesnumbered,algoruled,boxed,lined]{algorithm2e}
\usepackage[dvipsnames]{xcolor}
\usepackage{multirow}
\usepackage[backend=biber,sorting=none]{biblatex}

\DeclareMathOperator*{\argmin}{argmin}

\newtheorem{thm}{\textbf{Theorem}}
\newtheorem{propn}{\textbf{Proposition}}

\newtheorem{lemma}{\textbf{Lemma}}

\newcommand{\makeblack}{\color{black}}
\newcommand{\makered}{\color{red}}

\begin{document}
%
\title{Multi-Resolution Data Fusion \\ for Super Resolution Imaging}
%
%
%

\author{Emma~J.~Reid,~\IEEEmembership{Student Member,~IEEE,}
        Lawrence~F.~Drummy,        Charles~A.~Bouman,~\IEEEmembership{Fellow,~IEEE}
        and Gregery~T.~Buzzard~\IEEEmembership{Senior Member,~IEEE}
        
\thanks{E. J. Reid  is with the Department of Mathematics, Purdue University, West Lafayette, IN 47907, USA (e-mail: reid59@purdue.edu). }
\thanks{L. F. Drummy is with the Materials and Manufacturing Directorate, Air Force Research Laboratory, Wright-Patterson Air Force Base, OH 45433, USA (e-mail: lawrence.drummy.1@us.af.mil)}
\thanks{C. A. Bouman is with the Departments of Electrical and Computer
Engineering, and Biomedical Engineering at Purdue University, IN 47907, USA  (e-mail: bouman@purdue.edu).}
\thanks{G. T. Buzzard is with the Department of Mathematics, Purdue University, West Lafayette, IN 47907, USA  (e-mail: buzzard@purdue.edu).}%
\thanks{Manuscript received October 19, 2021;}
}

%
%

\markboth{Journal of \LaTeX\ Class Files,~Vol.~14, No.~8, August~2015}%
{Shell \MakeLowercase{\textit{et al.}}: }
%



\maketitle

\begin{abstract}
    Applications in materials and biological imaging are limited by the ability to collect high-resolution data over large areas in practical amounts of time.
    One solution to this problem is to collect low-resolution data and interpolate to produce a high-resolution image.  However, most existing super-resolution algorithms are designed for natural images, often require aligned pairing of high and low-resolution training data, and may not directly incorporate a model of the imaging sensor.

    In this paper, we present a Multi-resolution Data Fusion (MDF) algorithm for accurate interpolation of low-resolution 
    electron microscope 
    data at multiple resolutions up to 8x. Our approach uses small quantities of unpaired high-resolution data to train a neural network prior model denoiser and then uses the Multi-Agent Consensus Equilibrium (MACE) problem formulation to balance this denoiser with a forward model agent that promotes fidelity to measured data.
    
    A key theoretical novelty is the analysis of mismatched back-projectors, which modify typical forward model updates for computational efficiency or improved image quality. We use MACE to prove that using a mismatched back-projector is equivalent to using a standard back-projector and an appropriately modified prior model.
    We present electron microscopy results at 4x and 8x interpolation factors that exhibit reduced artifacts relative to existing methods while maintaining fidelity to acquired data and accurately resolving sub-pixel-scale features. 
    \makeblack
    \end{abstract}

\begin{IEEEkeywords}
Data Fusion, MACE, Multi-Agent Consensus Equilibrium, MDF, Multi-Resolution Data Fusion, Super Resolution, Plug-and-Play
\end{IEEEkeywords}

%
\IEEEpeerreviewmaketitle

%
%
%
%
\section{Introduction}
\label{sec:introduction}

\IEEEPARstart{M}{any} important material science problems require the collection of high resolution (HR) data over large fields of view (FoV).
For example, high resolution images are needed to extract detailed features, such as the 4 nanometer curli fibers structures in \textit{E.~coli}, which are fundamental in the formation of bacterial biofilms, or the 10-20 nm structures in gold nanorods materials, which are of interest due to their near-infrared light tunability and biological inertness \cite{park_nanorods}. 
Also, a large FoV is typically required to collect the representative volumes (RV) of materials that are needed to determine macroscopic properties such as material toughness and fracture strength \cite{he_construction_2021}.  

However, imaging multiple, large FoVs at high resolution is difficult under realistic constraints.
For example, raster scanning a $1\text{mm} \times 1\text{mm}$ FoV at a resolution of 10nm requires the acquisition of approximately $10$ giga-pixels of data, which requires roughly 17 hours under conditions described in \cite{anderson_sparse_2013}.  

One approach to overcoming this barrier is to acquire a large FoV at low resolution (LR) and interpolate it to obtain a higher resolution image of sufficient quality.  Ideally, 4x interpolation in each direction leads to a 16x decrease in acquisition time, while 8x interpolation leads to a 64x decrease.  

Traditional interpolation methods such as splines \cite{splines} do not offer sufficient quality, but recent advances using deep neural networks (DNNs) have produced a number of methods for high quality interpolation of natural images. For example,  \cite{dong_learning_2014} and \cite{yamanaka_fast_2018} use end-to-end DNNs trained on HR/LR paired images.  
This approach is improved in SRGAN \cite{ledig_photo-realistic_2017} through adversarial training and perceptual loss.  
Another approach is EnhanceNet \cite{sajjadi_enhancenet_2017}, which uses automated texture synthesis and perceptual loss but focuses on creating realistic textures rather than reproducing  ground truth images.  ESRGAN \cite{wang2018esrgan} introduces architectural improvements to SRGAN, and ESRGAN+ \cite{rakotonirina_esrgan_2020} adds further refinements.  
DPSR \cite{zhang_deep_2019} uses a form of Plug-and-Play (PnP) as described later, but still requires a CNN super-resolver trained on HR/LR pairs.  
And finally, DPSRGAN \cite{zhang_deep_2019} attempts to fuse these Plug-and-Play methods with the realistic textures generated by GANs.

\begin{figure*}
  \centering
  \includegraphics[width = 4.5in]{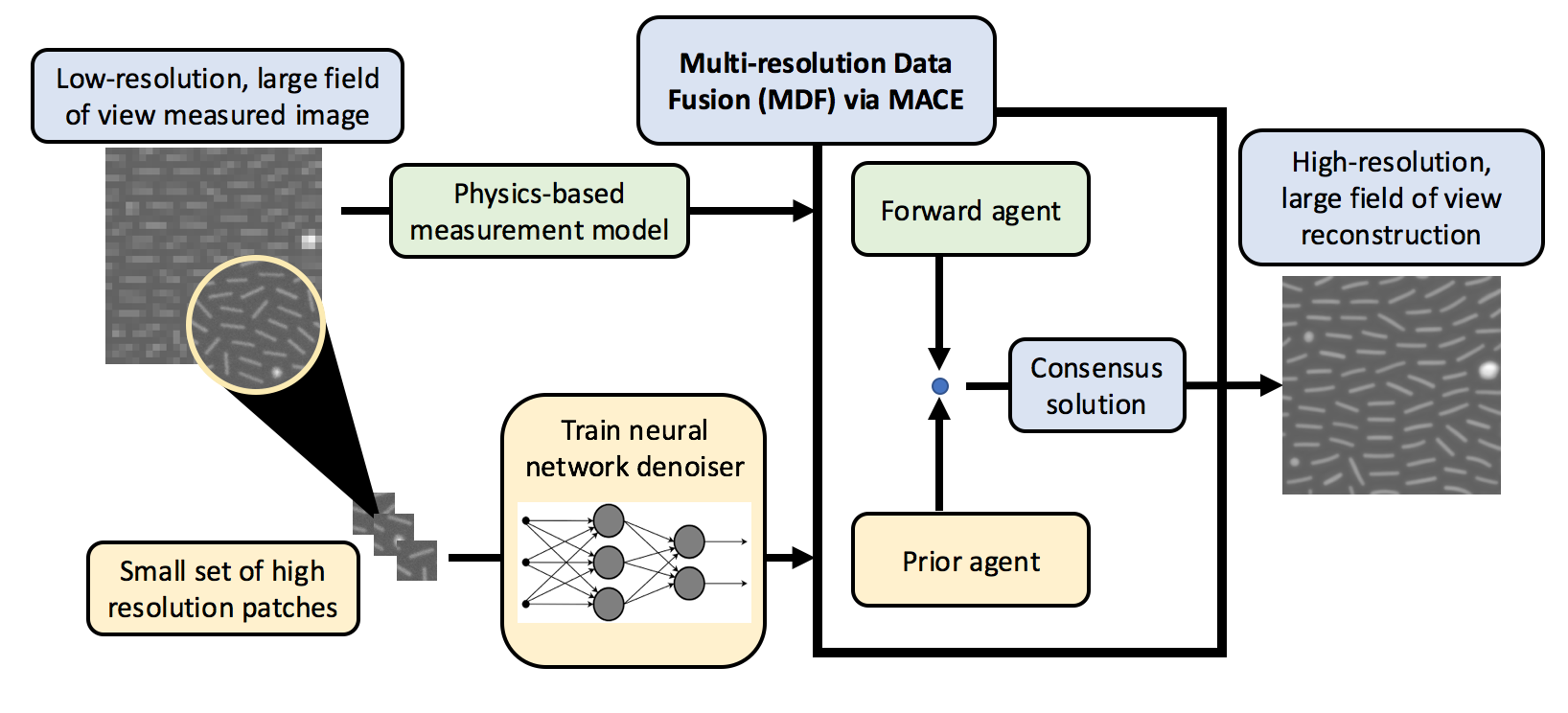}
  \caption{Overview of the MDF pipeline.  A small set of high-resolution data (unpaired with low-resolution data) is used to tune a CNN denoiser.  This tuned denoiser is used in the MACE framework with a microscopy-based forward model to produce high-resolution images from  low-resolution data.  }
  \label{fig:overview}
\end{figure*}

An impediment to applying these DNN methods to material imaging problems is that they are typically trained using a large set of aligned HR/LR patch pairs.
However, in practice it is difficult to acquire large quantities of accurately aligned HR/LR data \cite{Qian_2020}.
More recently, zero-shot learning has been proposed as a method that does not require aligned HR/LR data for training.
In zero-shot learning, the LR image is further downsampled to produce paired data to train a small, image-specific CNN that is used to upsample the original \cite{shocher_zero-shot_2018}. 
While this allows for quick and accurate reconstructions, it does not make use of high-resolution data in its training.

In this paper, we update the MDF algorithm of \cite{sreehari_multi-resolution_2017} to use a prior model based on deep learning and give a theoretical justification for the use of mismatched backprojectors.  Our method yields \makeblack accurate 4x and 8x interpolation of large FoV low-resolution EM images using selected unpaired patches of HR data. 
Moreover, our method can perform interpolation by a range of factors without retraining of the prior model.

More specifically, the novel contributions of this work include:
\begin{itemize}
        \item  An update of the MDF algorithm from \cite{sreehari_multi-resolution_2017} to incorporate recent advances in deep learning and a description of this algorithm in the MACE problem formulation. 
        \item The use of the MACE formulation to show that the use of a mismatched or relaxed adjoint projector (RAP) is provably equivalent to using the standard back projector with an appropriately modified prior. See Section III for more details on RAP. \makeblack
        \item Experimental results indicating that interpolation factors of 4x to 8x are possible with realistic transmission and scanning electron microscopy data sets.
\end{itemize}

 Our MDF-RAP approach can be applied to a number of imaging processing tasks and modalities, and its modularity allows for changes in forward models, resolution level, and regularization strength without retraining.  Our analysis of the Relaxed Adjoint Projection interprets this particular change to the forward model to a corresponding change to the prior.
\makeblack

In Figure \ref{fig:overview}, we provide a visual representation of the MDF approach.
Using a LR base image, we train a denoiser on a sparse set of HR patches of the same (or similar) specimen.
This allows the denoiser to learn the underlying manifold of the HR data while simultaneously being trained to remove additive white Gaussian noise (AWGN).
This denoiser is then applied in the MACE framework to achieve a super-resolution reconstruction of the low-resolution base image. 
This approach does not require registered pairs of HR and LR data, allowing for flexible levels of super resolution and simple generalization to other problems. 
Additionally it allows for use of known forward models and has a single parameter that can be used to control the weight of data fidelity relative to strength of regularization.  

Code for this project is available at https://www.github.com/emmajreid/MDF.

\section{Related Work \& Problem Formulation}
 Data fusion, or the combination of multiple sources of data, has been used successfully in domains such as medical imaging, remote sensing, and others for many years  \cite{multimodalMR, moselyAHA}. 
In medical imaging, scientists combine scans from multiple sources such as PET, CT, and MRI to reconstruct images that better show internal body properties \cite{ALAM201924}. 
Recent work on data fusion in remote sensing \cite{hongSAR} provides a framework for multi-modal and cross-modal deep learning methods for pixel-wise classification and spatial information modeling.  In a separate direction, \cite{gaoSR} applies super-resolution in the spectral domain to images that are spatially high-resolution but spectrally low-resolution; this is done by learning a joint sparse and low-rank dictionary of partially overlapping hyperspectral and multispectral images and their sparse representations.  In \cite{yao2020cross_attention} this is done with an unsupervised hyperspectral super-resolution network that blends spatial information from multispectral sensors and spectral information from hyperspectral sensors.  

The PnP algorithm \cite{venkatakrishnan_plug-and-play_2013, sreehari2016plug} and MACE problem formulation \cite{buzzard_plug-and-play_2018} provide a framework in which multiple models can be combined with little modification, thus providing a natural approach to data fusion.   
A PnP approach was used in \cite{ghani_data_2020} to combine CT and MRI modalities across sensor, data, and image priors, but all at a single resolution. 
In \cite{7532589}, PnP was used with the Nonlocally Centralized Sparse Representation algorithm to combine sparse coding and dictionary learning to turn a denoiser into a super-resolver. 
However this method begins to break down in the presence of noise, which is prevalent in microscopy images.

In work closely related to the current paper, \cite{sreehari_multi-resolution_2017} uses PnP to develop MDF for microscopy.  In that work, they used a fixed library of domain-specific HR patches as the basis for a non-local means denoiser and then coupled this with a data-fitting forward model to achieve super-resolution from LR image data over a large field of view.  However, the resulting denoiser is computationally expensive relative to a neural network approach.  Moreover, that paper did not provide theoretical foundation to guide modifications to the algorithm.  

As noted above, we extend the work in \cite{sreehari_multi-resolution_2017} to use a neural network prior and describe our method and the use of RAP in terms of the MACE formulation.  
In Section~\ref{microscopy} to \ref{mace}, we examine the modeling of the microscope's point spread function, detail the forward model derivation from \cite{sreehari_multi-resolution_2017}, motivate RAP, and describe the MACE framework developed in \cite{buzzard_plug-and-play_2018} and the relationship between MACE and PnP.

\makeblack

\subsection{Microscopy Forward Model}\label{microscopy}
Our goal is to interpolate a rasterized image $y \in \mathbb{R}^M$ to a HR version $x \in \mathbb{R}^N$.  
Super-resolution by a factor of $L$ implies scaling by $L$ in each direction, so that $N = L^2M$.  For such a problem, the forward model is typically
\begin{equation}\label{eq:forwardmodel}
    y = \Psi x + \epsilon,
\end{equation}
where $\Psi \in \mathbb{R}^{M\times N}$ represents the point spread function of the microscope and $\epsilon \sim N(0, \sigma_w^2)$ is an $M$ dimensional vector of AWGN. For our application, $\psi$ is closely modeled by averaging over $L \times L$ patches. \makeblack
The data-fitting cost function is then $\frac{1}{2} \|y - \Psi x\|^2$, which is embedded in a proximal map and balanced by the action of a prior model.  For application purposes, we now need to approximate $\Psi$.

In bright-field transmission electron microscopy, a parallel beam of electrons illuminates a thin material sample, and the resulting transmitted beam is formed into an image using the microscope’s objective lens. 
This is then further magnified using the microscope’s projection lens system.
Using this configuration, the transmission electron microscope can yield a magnification ranging from 1,000X to over 1,000,000X. 
The magnified beam is sampled at the image plane using a pixel array detector such as a direct electron detector or CCD camera. 
Since the electron-optical magnification can be controlled and the image is detected using a pixel array detector, a block-averaging forward model is a good approximation for this acquisition modality.  
In this forward model, a square region in the high-resolution image is averaged to produce a single pixel value in the low-resolution image.  

\begin{figure}
  \includegraphics[width = 3.5in]{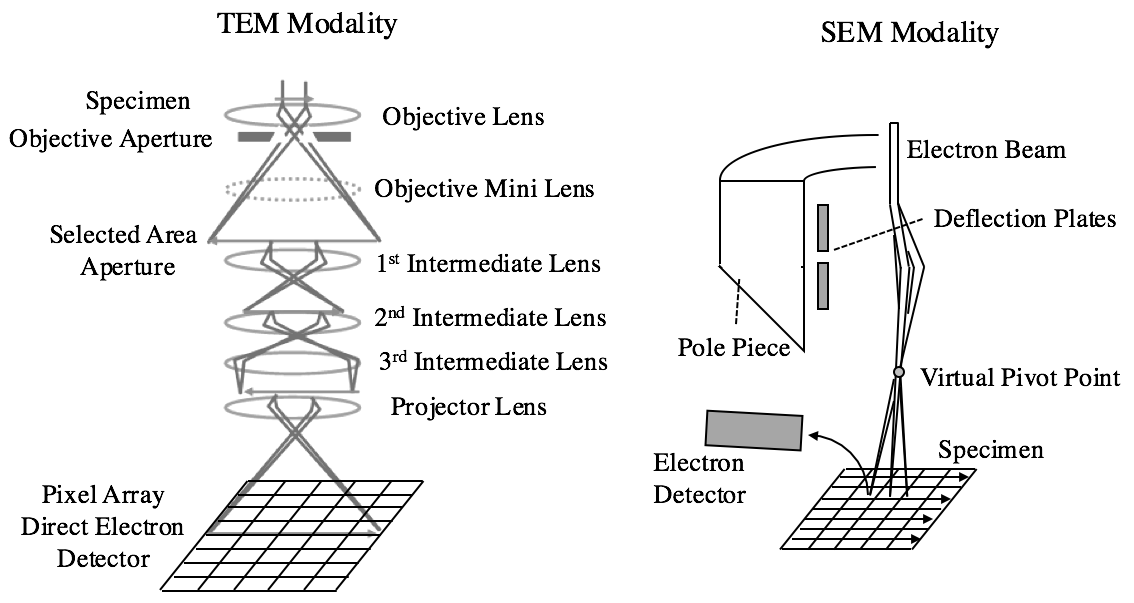}
  \caption{Illustration of the transmission and scanning electron microscopy image acquisition process. We approximate both image acquisition processes by block-averaging a HR image as described in Section~\ref{pnpformulation}. The scanning electron microscopy image shown is adapted from \cite{bree-SEM}.}
  \label{fig:microscopy}
\end{figure}

For scanning electron microscopy, an electron beam is focused to a small diameter probe which is then raster scanned across the surface of the sample using beam deflectors. 
As the probe strikes the sample, secondary electrons are ejected from the sample and collected with an integrating detector, which sums the total number of electrons scattered at each point on the surface of the sample. 
The raster array dimensions can be controlled to give LR and HR data, so again a block-averaging forward model is a good approximation to the imaging system. 
The image acquisition process for transmission and scanning electron microscopy is depicted in Figure \ref{fig:microscopy}.

\subsection{PnP Formulation}{\label{pnpformulation}}
Plug-and-play (PnP) \cite{venkatakrishnan_plug-and-play_2013, sreehari2016plug} is an algorithm for solving inverse problems that replaces the probabilistic encoding of Bayesian prior information with an algorithmic encoding; a candidate reconstruction is denoised to produce a more plausible reconstruction without using a cost function or likelihood. This prior update is combined iteratively with a forward update until a fixed point is reached.  
Using the description in Section~\ref{microscopy}, we approximate the forward model $\Psi$ by block-averaging every non-overlapping neighborhood of $L\times L$ pixels in the HR image to obtain the LR image \cite{sreehari_multi-resolution_2017}. 
For notational convenience, we let $A$ represent summation over $L \times L$ blocks, in which case $\Psi \approx A / L^2$ in \eqref{eq:forwardmodel}.  
Also, $A^T$ is given by replicating each pixel into an $L \times L$ block, so $A A^T = L^2 I$.  The negative log-likelihood is then 

\begin{equation}
 l(x) = \frac{1} {2\sigma_w^2} \left \|y-\frac{1}{L^2} A x\right \|^2+\frac{M} {2}\log(2\pi \sigma_w^2).
\end{equation}

As part of the Plug-and-Play algorithm, Sreehari et al.~used the proximal map of $l(x)$, given by
\begin{equation} \label{eq:1storderopt}
  F(x; \sigma_\lambda) = \underset{\hat{x} \geq 0}{\mathrm{argmin}} \left[ l(\hat{x}) + \frac{1} {2 \sigma_\lambda^2} ||x-\hat{x}||^2_2\right].  
\end{equation}
%

As in the proof of Lemma 1 in the Appendix, we use the fact that $A A^T = L^2 I$ to reparameterize the solution of \eqref{eq:1storderopt} in a simpler form.  We also use clipping to impose the physical constraint of nonnegativity as in the constrained minimization of \eqref{eq:1storderopt}.  This yields
\makeblack 
%
\begin{equation}\label{eq:5}
    F(x; \sigma_\lambda) = \left[x + \frac{\sigma_\lambda^2}{\sigma_\lambda^2 + L^2 \sigma_w^2} A^T\left(y-\frac{1} {L^2}Ax\right)\right]_+.
\end{equation}

The block replication inherent in $A^T$ can lead to blocky artifacts in the final reconstruction, which motivates our introduction of the Relaxed Adjoint Projector in Section~\ref{Multi-Resolution Data Fusion}. 
Sreehari et al. used a library-based Non-Local Means (NLM) prior model to incorporate HR data, which led to slow reconstruction times due to the computational cost of NLM \cite{nlmissues}. For this reason, we opted to use a deep neural network denoiser as our prior model.
PnP has been used with similar denoisers as prior models in \cite{zhang2017learning} \cite{zhang2020plug} \cite{SHI2020}. 
However, limitations of this previous work include little ability to control regularization strength and the use of generically trained neural networks as opposed to domain-specific denoisers.


\subsection{MACE Reconstruction Framework}\label{mace}

In PnP, which is derived from the ADMM solution to a Bayesian inversion problem, forward and prior models can be designed separately, thus promoting modularity and separating sensor modeling from image domain modeling.  However, the resulting fixed point of the PnP algorithm no longer has the probabilistic interpretation given in the Bayesian approach.

Multi-Agent Consensus Equilibrium (MACE) \cite{buzzard_plug-and-play_2018} provides a formulation and interpretation of the problem solved by the PnP algorithm and leads naturally to generalizations and analysis of PnP.  MACE describes the fixed point of PnP in terms of an equilibrium condition and extends the algorithmic denoisers of PnP to a framework for the fusion of multiple forward and prior models;  MACE also provides parametric control of regularization. 
MACE itself is not an algorithm, but the solution of the MACE equations can be found with PnP or with other algorithms for solving systems of equations. 

\makeblack


The motivating problem for MACE is to minimize the function
\begin{equation} \label{eq:6}
f(x)=\sum \limits_{i=1}^{K} \mu_i f_i(x_i) \text{ s.t. } x_i = x, \: i = 1, \ldots, K,
\end{equation}
with $x, x_i \in \mathbb{R}^N$ and weights $\mu_i >0$ with $\sum \limits_{i=1}^{K} \mu_i = 1$.  

Buzzard et al. \cite{buzzard_plug-and-play_2018} describe a framework that generalizes \eqref{eq:6} to include algorithmic priors and forward maps (sometimes called agents).
For $K$ vector valued maps, $F_i: \mathbb{R}^N \to \mathbb{R}^N$, $i=1, \ldots, K$, define the consensus equilibrium for these maps to be any solution $(x^*, \textbf{u}^*) \in \mathbb{R}^N \times \mathbb{R}^{NK}$ such that 
\begin{equation}\label{eq:2}
    F_i(x^*+\textbf{u}_i^*) = x^*, \: i=1, \ldots K  \ \ \text{ and } \ \ 
    \overline{\textbf{u}}_\mu^* = 0
\end{equation}
where \textbf{u} is a vector in $\mathbb{R}^{NK}$ constructed by stacking vectors $u_1, \ldots u_K$, and $\overline{\textbf{u}}_\mu$ is the weighted average $\overline{\textbf{u}}_\mu = \sum \limits_{i=1}^{K} \mu_i u_i$.
In the case of \eqref{eq:6}, the maps $F_i$ are chosen to be proximal maps associated with the $f_i$.

The conditions in \eqref{eq:2} are equivalent to a related system of equations.  
Namely for $\textbf{v} \in \mathbb{R}^{NK}$ with $\textbf{v} = (v_1^T, \ldots, v_K^T)$, $v_i \in \mathbb{R}^N \: \forall i$,  define $\textbf{F}(\textbf{v})$ by stacking the vectors $F_j(v_j)$, and define $\textbf{G}_\mu(\textbf{v})$ by stacking $K$ copies of $\overline{\textbf{v}}_\mu$. Additionally for $x \in \mathbb{R}^N$, define $\hat{x}$ to be the vector obtained by stacking $N$ copies of $x$.
Then $(x^*, \textbf{u}^*)$ satisfies \eqref{eq:2} if and only if $\textbf{v}^* = \hat{x}^*+\textbf{u}^*$ satisfies $\overline{\textbf{v}}_\mu^* = x^*$ and 
\begin{equation} \label{eq:FG}
    \textbf{F}(\textbf{v}^*) = \textbf{G}_\mu(\textbf{v}^*).
\end{equation}
As in PnP, the $F_i$ may be replaced by more general operators such as CNN denoisers, in which case the solution of \eqref{eq:FG} is the fixed point of a generalized PnP algorithm.  This is solved with Mann iterations in 
 \cite{majee_4d_2019} as shown in Algorithm 1, which we use for our results.   
\begin{algorithm}
\SetAlgoLined
 \textbf{Input:} Initial Reconstruction $x^{(0)} \in \mathbb{R}^N$ \\
 \textbf{Output:} Final Reconstruction $x^*$\\
 \textbf{x} $\leftarrow$ \textbf{v} $\leftarrow$ $\begin{pmatrix} x^{(0)}\\ \vdots \\ x^{(0)} \\ \end{pmatrix}$ \\
 \While{unconverged}{
  $\textbf{x} \leftarrow \textbf{F}(\textbf{v})$ \\
  $\textbf{z} \leftarrow \textbf{G}_\mu(2\textbf{x}-\textbf{v})$ \\
  $\textbf{v} \leftarrow \textbf{v}+2\rho (\textbf{z}-\textbf{x})$\\
 }
 $x^* \leftarrow \overline{\textbf{v}}_\mu$
 \caption{Algorithm to find MACE solutions}
\end{algorithm}

In the case of 2 agents as in our results below, this algorithm is equivalent to the PnP algorithm of \cite{sreehari_multi-resolution_2017}, but the MACE formulation is central to the analysis of RAP presented below.

\section{RAP and MDF} 
\label{Multi-Resolution Data Fusion}

A key strength of the MACE framework is the ability to incorporate algorithmic denoisers and other operators.  
In this section, we leverage this observation in two ways.

First, we show that using a mismatched backprojector in \eqref{eq:5} is equivalent to using the original forward operator in \eqref{eq:5} with a modified prior agent.  We call the mismatched backprojector a Relaxed Adjoint Projector (RAP).  More precisely, we modify the forward operator in \eqref{eq:5} by replacing $A^T$ with a related matrix $B$. 
This change means that the forward operator is no longer the proximal map for $l(x)$.  
However, we show using the MACE framework that this formulation is equivalent to a formulation using the original forward operator in \eqref{eq:5} but with an alternative prior that depends on $B$.  
This provides important intuition for the use of RAP and allows for the application of existing convergence results.  

Second, we describe our method of MDF, in which representative HR patches are collected independently of the LR scan. 
These patches are then used to train a denoising prior model, which intrinsically learns the underlying distribution of the high-resolution modality. 
We then perform LR scans over a large area and fuse the two modalities using the Multi-Agent Consensus Equilibrium framework to produce a HR image encompassing the full FoV.  

\subsection{Relaxed Adjoint Projection} \label{forward}
 The matrix $A^T$ in \eqref{eq:5} is often called a back-projector because it takes the error between measurements $y$ and the forward model and projects them back to image space.  In practice, a different matrix may be used in place of $A^T$; this is known in some contexts as using a mismatched backprojector.  This may be done for computational simplicity, as in \cite{venkatBP}, or because the mismatch leads to improved results, as in earlier work on iterative filtered backprojection \cite{lalush_improving_1994} \cite{ziabari_mismatch_2018} \cite{schofield_image_2020}.
\makeblack 

 In the context of \eqref{eq:5}, we provide an interpretation of this mismatch, which we term Relaxed Adjoint Projection (RAP), as equivalent to the standard backprojector with an appropriately modified prior.  That is, the equilibrium problem obtained with a given prior update map and a forward update with a mismatched backprojector has the same solutions as a modified prior map and the standard backprojector update.
\makeblack 

For MDF-RAP, we consider the RAP update map
\begin{equation} \label{eq:mismatch}
    \tilde{F}(x; \sigma_\lambda) = \left [x + \frac{\sigma_\lambda^2}{\sigma_\lambda^2 + L^2 \sigma_w^2} B\left (y-\frac{1} {L^2}Ax \right) \right]_+, 
\end{equation}
where the matrix $B$ represents bicubic upsampling by a factor of $L$ and replaces the block replication operator $A^T$.  
We note that while bicubic backprojection is used here to avoid blocky artifacts observed when using $A^T$, other interpolants can be used.

We first describe the equilibrium problem associated with RAP, then prove that the solution of this problem arises from 3 different formulations:
\begin{itemize}
    \item {\bf RAP forward update}, standard prior, equal weight averaging
    \item standard forward update, standard prior, {\bf modified averaging}
    \item standard forward update, {\bf modified prior}, equal weight averaging.
\end{itemize}
Thus, replacing $A^T$ with $B$ is equivalent to using $A^T$ with a modified prior that can be described in terms of $B$.

In general, the update in \eqref{eq:mismatch} is not a proximal map for any function when the matrix $BA$ is not symmetric since a symmetric Jacobian is a property of all proximal maps.
By converting the RAP update into a modified prior, we recover the ability to use standard convergence results while maintaining the benefits associated with mismatched backprojection.  
We note that results in \cite{chouzenoux:hal-02961431} prove convergence for an adjoint mismatch in the Proximal Gradient Algorithm but do not address the equivalence described here. 
 
To describe RAP further, note that the first-order optimality condition for a solution of \eqref{eq:6} when all $\mu_j$ are equal is 
\begin{equation}
   \nabla f_1(x^*) + \cdots + \nabla f_K(x^*) = 0.
\end{equation}
Also, the proximal map for a convex and differentiable $f_j$ is given by $F_j(v_j) = x_j = v_j - \nabla f_j(x_j)$; i.e, the update can be regarded as an implicit gradient descent step, with the gradient evaluated at the output point $F_j(v_j) = x_j$.  

In the case of a mismatched backprojector, we assume for the moment that $F^{R_j}_j$ is given by $F^{R_j}_j(v_j) = x_j = v_j - R_j \nabla f_j(x_j)$ for some matrix $R_j$, which we think of as close to the identity (precise conditions on $R$ are given in the theorem statements in the appendix).
In the case of \eqref{eq:5} and \eqref{eq:mismatch}, this corresponds to $R A^T = B$, and if $B$ is close to $A^T$, then $R$ can be chosen to be close to the identity. 
\makeblack
Using this $F^R$ and all $\mu = 1/K$ in the equilibrium condition \eqref{eq:FG}, we have
\begin{equation} \label{eq:equilibrium}
    v_j^* - R_j \nabla f_j(x_j^*) = \frac{1}{K} \sum_k v_k^*.
\end{equation}
Since the left hand side is $x_j^*$ for each $j$ and the right hand side is independent of $j$, we have $x_j^* = x^*$ is independent of $j$.  Summing these equations over $j$ and subtracting the sum of the $v_j^*$ from both sides and taking the negative gives
\begin{equation}\label{eq:15}
    R_1 \nabla f_1(x^*) + \cdots + R_K \nabla f_K(x^*) = 0.  
\end{equation}
This is the corresponding equilibrium condition for the operators $F^{R_j}_j$ and equal weight averaging.  

Theorem~1 states that the set of solutions of the equilibrium condition with RAP are the same as those obtained using standard back projections but an alternative averaging operator $\textbf{G}^R$, given by a matrix-weighted average using the $R_j$ as matrix weights. 
As before, we stack the operators $\textbf{F}_j^{R_j}$ to obtain $\textbf{F}^R$.   The details of the notation, hypotheses, and the proof are given in the appendix. 

\vspace{0.1in}
{\bf \em Theorem} {\em 1:}
Under appropriate hypotheses (see appendix) on $\textbf{F}$ and $R$, there is a map from solutions  $\textbf{v}^*$ of 
$$\textbf{F}^R(\textbf{v}^*) = \textbf{G}(\textbf{v}^*)$$
to solutions $\textbf{\^ v}^*$ of 
$$\textbf{F}(\textbf{\^ v}^*) = \textbf{G}^R(\textbf{\^ v}^*)$$
such that for each such pair, $\textbf{G}(\textbf{v}^*) = \textbf{G}^R(\textbf{\^ v}^*)$.  There is also such a map from $\textbf{\^ v}^*$ to $\textbf{v}^*$.  
\vspace{0.1in}

The map $\textbf{G}(\textbf{v}^*)$ is obtained by stacking the solution $x^*$, so this theorem implies that these two formulations have the same set of possible reconstructions.     
The following corollary applies this to give the equivalence between the use of mismatched backprojection in the data-fitting operator and the use of standard backprojection with an alternative prior.  

\vspace{0.1in}
{\bf \em Theorem} {\em 2:} With appropriate assumptions (see appendix) on $B = RA^T$ and the denoiser $H$, and with equal weighting $\mu_j = 1/2$, the following two choices lead to the same MACE solution in \eqref{eq:FG}:
\begin{itemize}
    \item $F_1 = \tilde{F}$ is the RAP update in \eqref{eq:mismatch} and $F_2 = H$ is a given prior operator;
    \item $F_1 = F$ is the standard update in \eqref{eq:5} and $F_2 = H \circ \Phi_R$, where $\Phi_R$ is a function dependent on the matrix $R$.  
\end{itemize} 

This makes precise the idea that the mismatch in RAP is equivalent to a corresponding modification to the update step of the prior operator.  
By using the mismatch, which is often more efficiently implemented in the form of RAP than with alternative averaging, we gain the ability to more closely match the prior to the observed structure of the data without changing the algorithmic implementation of the prior.  

\subsection{Convergence of Relaxed Adjoint Projection (RAP)}

From \cite{buzzard_plug-and-play_2018}, Algorithm 1 is known to converge when the map $\textbf{v} \mapsto 2\textbf{F}(\textbf{v})-\textbf{v}$ is nonexpansive, and this condition is satisfied when each $F_j$ is the proximal map for a convex function.  
When the RAP update is used, then $F_j$ is typically no longer a proximal map and may not be nonexpansive.  

In Theorem~3 we give conditions under which $F_j$ using RAP is a proximal map after an appropriate change of coordinates.  
The key idea, related to work in \cite{nairconv}, is to consider operators of the form $F(x) = Wx + q$.  
When $F$ is a proximal map, then $W$ is symmetric and positive-definite.  A change of variables allows us to recover this property even for some non-symmetric $W$. 
This allows us to prove convergence of Algorithm~1 when using the RAP forward model and denoiser $H$ as in Theorem 2.  The proof is in the appendix.    

\vspace{0.1in}
{\bf \em Theorem} {\em 3:}  
Let $F(x) = Wx+q$ where $W = V \Lambda V^{-1}$ is an $N \times N$ matrix, $\Lambda$ is diagonal with eigenvalues in $(0,1]$, and $q \in \mathbb{R}^N$, and let $H$ be a denoiser such that  $V^{-1}H V$ is nonexpansive.
Then the Algorithm~1 converges using the operators $F_1=F$ and $F_2=H$. 
\vspace{0.1in}

\subsection{Multi-Resolution Data Fusion} \label{denoisers}

The MACE formulation with either the standard or RAP forward operator provides a natural way to incorporate low-resolution data and maintain fidelity of the reconstruction to this data.  
For MDF, we use a neural network denoiser as a prior agent to incorporate selected high-resolution data, either from the image under reconstruction or from related images. 

The theoretical foundation of Plug-and-Play implies that the prior operator should be a denoiser for images in the target distribution perturbed by AWGN, in principle independent of the noise present in the data itself \cite{bouman_plug-and-play_2021}.  
However, as seen in \cite{venkatakrishnan_plug-and-play_2013} the prior operator can play a large role in the quality of the final reconstruction.  
In the context of learned denoisers such as CNNs, this means that the CNN must denoise well on the images in the distribution under consideration.  
Since we use an iterative algorithm, the CNN must also denoise well on neighboring images in order to converge to a high-quality reconstruction.  
Ideally, the denoiser should be able to take any image in the reconstruction space and move it closer to an image in the target distribution, but in practice we must settle for an approximation based on a sparsely sampled set of images near the distribution.  

We note here that increasing the super-resolution factor $L$ necessarily increases the set of data-consistent reconstructions -- increasing $L$ increases the dimension of the kernel of $A$.  
In particular, given two reconstructions that both fit the data equally well and that are both equally well-denoised by the denoiser (more precisely, both are equilibrium solutions), there is no reason to favor one over the other.  
This means first that the importance of the prior operator increases with $L$ and second that larger $L$ gives any reconstruction algorithm more opportunity to ``hallucinate'' detail that may or may not be present in the true image.  
In the context of scientific and medical applications where the reconstruction can influence important decisions, it can be detrimental to push the limits of super-resolution and/or regularization beyond reasonable expectations \cite{antun_instabilities_2020}.  

\section{Methods and Results} \label{Results}

We apply the MDF-RAP method on 3 microscopy datasets with pronounced differences in data distribution:  gold nanorods, an \textit{E. coli} biofilm, and pentacene crystals.  
The gold nanorod images are composed of non-overlapping, nearly linear segments at various angles with nearly circular impurities.   
The \textit{E. coli} biofilm images contain a wide variety of shapes and textures as well as large regions of nearly empty space.   The pentacene crystal images are typically composed of large regions of relatively constant intensity with sharply defined edges.
The nanorod and pentacene images were obtained using scanning electron microscopy, while the \textit{E. coli} images were obtained using transmission electron microscopy.

\makeblack
We present comparisons of MDF-RAP with a variety of alternatives for both 4x and 8x interpolation.
At 4x we compare with bicubic interpolation, DPSR, DPSRGAN \cite{zhang_deep_2019}, and PnP with DnCNN trained on natural images.
However, at 8x, we do not compare with DPSRGAN since it is not available for this interpolation rate.

\subsection{Computational Methods}

We use a MACE formulation as in equation \eqref{eq:FG} with two agents, one for data fidelity and one to incorporate prior information in the form of a denoiser. With 2 agents, the relative weights $\mu_i$ simplify to one weight $\mu$ for the forward agent and the complementary weight $1 - \mu$ for the prior agent.  Since the prior agent operates in the reconstruction space of HR images, we train a CNN denoiser to remove 10\%  AWGN from high-resolution target images.  

We use Algorithm 1 to determine the corresponding reconstructions. Based on the MACE equation $\textbf{F}(\textbf{v}) = \textbf{G}(\textbf{v})$, we define a measure of convergence error as
\begin{equation} \label{eq:conv-error}
    \text{Convergence Error} = \frac{||\textbf{G}(\textbf{v}) - \textbf{F}(\textbf{v})||_2} {\sigma_n ||\textbf{G}(\textbf{v})||_2},
\end{equation}
where $\sigma_n$ is the noise level used to train the prior model.   We display a plot of convergence behavior in Section~\ref{Synthetic}.

For a baseline case (labeled PnP below), we use the standard backprojector as in equation~\eqref{eq:5} and a DnCNN denoiser \cite{zhang_beyond_2017} trained on {\em natural images.}  We use $\mu=0.5$ for all results with this approach.  Algorithm 1 in this context is equivalent to Plug-and-Play \cite{buzzard_plug-and-play_2018}, hence the label PnP. This case is comparable to other methods with no domain-specific training.    

For the data fusion case (labeled MDF-RAP below), we use the RAP backprojector as in equation~\eqref{eq:mismatch} and a DnCNN denoiser trained on {\em representative high-resolution microscopy images.}  Although this involves two changes relative to the PnP baseline, Theorem~\ref{thm:RAP-equiv} implies that using RAP is equivalent to using the standard backprojector with a further modified prior.  In order to disambiguate RAP from the domain-specific prior, we illustrate the effect of RAP separately in Figure~\ref{fig:pentRAP}.  In the case of MDF-RAP, the use of RAP changes the effect of the prior, so we adjust the regularization strength using $\mu=0.8$ for synthetic data and $\mu=0.2$ for real data. 

\makeblack

The DnCNN architecture consists of 17 total layers with the following structure: (i) Conv+ReLU for the first layer with 64 filters of size 3x3x1. (ii) Conv+BatchNorm+ReLU for layers 2-16 with 64 filters of size 3x3x64. (iii) Conv for the last layer with 1 filter of size 3x3x64 \cite{zhang_beyond_2017}. 
The network uses a residual mapping $D$ to learn the structure of the noise in its training pairs $(x_\text{clean}, x_\text{noisy})$. 
A forward pass through the model is thus given by $\hat{x}_\text{clean} = x_\text{noisy} - D(x_\text{noisy})$.  We used code adapted from https://github.com/cszn/KAIR to implement DnCNN. 

To train each image-tuned prior for MDF-RAP, we randomly extracted 400 180x180 patches from a high-resolution training image. This represents a naive sampling of the high-resolution data. 
 We employed a 80/10/10 split for training, validation, and testing data. The amount of HR data used for training for each respective dataset is given in Table~\ref{tbl:speedup-results}. \makeblack
Using these training patches, we generate corresponding noisy patches by adding AWGN with standard deviation $\sigma = 0.1$. These patch pairs are then passed through the network for training (note that there is no pairing of high-resolution images with low-resolution images). 
We used an increase in validation loss as a stopping criterion to avoid overfitting.  
Each of our MDF-RAP networks trained for 1-2 hours using 1 Nvidia V100 GPU.

\subsection{Data Generation}

In order to provide quantitative accuracy metrics and demonstrate real-world behavior, we present two experimental approaches:  (i) partially simulated and (ii) fully real data.

\begin{figure*}
\vspace{0.1in}
\begin{center}
  \includegraphics[width = 1.in]{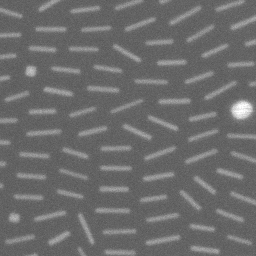}
  \includegraphics[width = 1in]{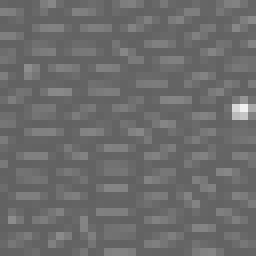}
  \includegraphics[width=1in]{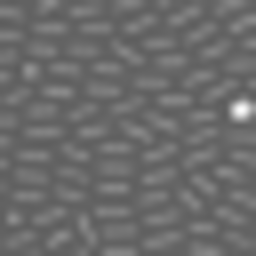}
  \includegraphics[width=1in]{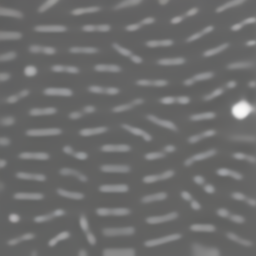}
  \includegraphics[width=1in]{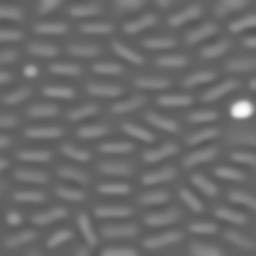}
  \includegraphics[width=1in]{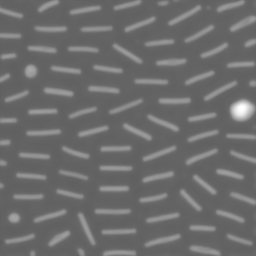}
\end{center}
\hspace{.7in} HR GT \hspace{.3in} Simulated LR \hspace{.4in} Bicubic \hspace{.5in} DPSR \hspace{.7in} PnP \hspace{.5in} \makered {\bf MDF-RAP} \makeblack 

\hspace{1.7 in} 26.27 dB  \hspace{.4in} 27.43 dB \hspace{.4in} 30.09 dB \hspace{.5in} 28.36 dB \hspace{.4in} \textbf{33.24 dB}

\caption{8x super-resolution reconstructions of a simulated LR EM image of gold nanorods with $\mu = 0.8$ for MDF-RAP. Each shows a field of view 569 nm wide.  MDF-RAP produces nanorods with clear edges and the proper shape, while each of the other methods includes significant blurring and/or incorrect shapes.}
\label{fig:8xNano}
\end{figure*}

\begin{figure*} 
\vspace{0.1in}
\begin{center}
  \includegraphics[width = 0.9in]{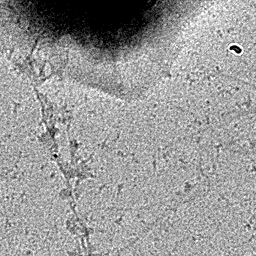}
  \includegraphics[width = 0.9in]{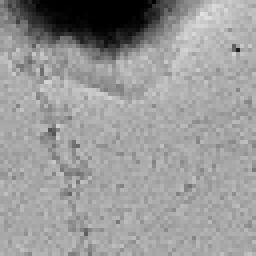}
  \includegraphics[width = 0.9in]{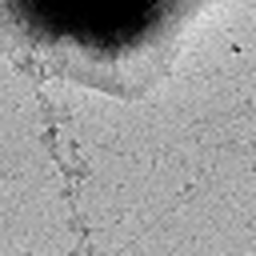}
  \includegraphics[width=0.9in]{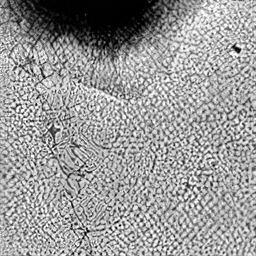}
  \includegraphics[width=0.9in]{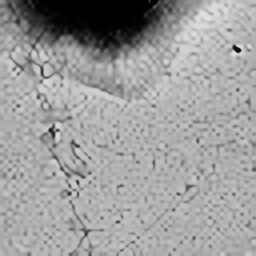}
  \includegraphics[width=0.9in]{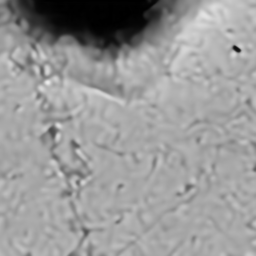}
  \includegraphics[width=0.9in]{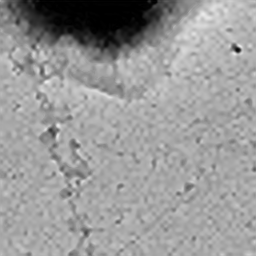}
\end{center}
\hspace{.45in} HR GT \hspace{.25in} Simulated LR \hspace{.25in} Bicubic \hspace{.3in} DPSRGAN \hspace{.35in} DPSR \hspace{.6in} PnP \hspace{.43in} \makered {\bf MDF-RAP} \makeblack

 \hspace{1.4in} 19.88 dB \hspace{.3in} 19.90 dB \hspace{.3in} 14.54 dB \hspace{.35in} 19.69 dB \hspace{.4in} 19.84 dB \hspace{.3in} \textbf{19.98 dB}

\caption{4x super-resolution reconstructions of a simulated LR EM image of E coli with $\mu = 0.8$ for MDF-RAP. Each shows a field of view 251 nm wide. MDF-RAP provides the best visual compromise between clarity and fit to data as shown by its reconstruction of distinct curli fibers and background preservation.}
\label{fig:4xecoli}
\end{figure*}

In the partially simulated case, we use actual HR microscopy data and then apply the forward model $x \mapsto Ax/L^2$ to obtain simulated LR data.  Each algorithm is applied to the LR data and the result is compared to the paired HR data using PSNR (and FRC in select cases) to provide quantitative accuracy measures.  

In the case of fully real data, both the HR and the LR data are obtained experimentally and used as in the previous case with the exception that we do not have aligned pairs and so cannot quantitatively compare the super-resolved output of an algorithm to a reference HR image.  These results on real data allow for qualitative assessment under practical application conditions.

For MDF-RAP, we use domain-specific HR images to train the CNN denoiser, but these are not required to be aligned with corresponding LR images, and the LR images are not used for training.  In all cases, we need LR data to illustrate super-resolution.  
All data used in testing the algorithm (both LR and HR) was taken from the test set and therefore excluded from MDF-RAP training data. 

Our HR data was collected using two separate microscopy methods to yield three data sets. Table~\ref{datasetinfo} describes the experimental parameters used for data acquisition.
\makeblack 
Scanning electron microscopy images for gold nanorods and pentacene were obtained on an FEI XL30 at 5kV with a secondary electron detector and a Zeiss Gemini at 5kV using an in-lens secondary electron detector. 
The interaction volume of the focused electron beam was on the order of the size of the resulting pixel size in the image. 
The Scanning Electron Microscopy modality raster scanned a focused beam across the sample with a pixel dwell time of 50 nanoseconds. 
TEM images for \textit{E. coli} were obtained on a 60-300 Thermo Fisher Titan operating at 300kV in bright field mode.
Images were collected on a 4k by 4k Gatan K2 Direct Electron Detector using Serial EM at various electron optical magnifications. 
LR overview images were first collected, followed by automated HR image montages. 
The bright field imaging modality uses a wide illumination that covers the entire imaging array.

\begin{table}[th] 
\caption{Acquisition parameters for experimental datasets. We omit LR pixel spacing for \textit{E. coli} as we do not perform super resolution on measured data in this case. The 2.2 nm nanorods sample was used for the results shown in Figure~\ref{fig:8xNano}.}
\begin{center}
\begin{tabular}{c|c|c|c} 
    Material & HR Pixel Spacing & LR Pixel Spacing & Data Modality\\
    \hline
    Nanorods & 1.1, 2.2 nm & 5.5 nm & Scanning\\
    \textit{E. coli} &  0.98 nm & N/A & Transmission\\
    Pentacene & 41.7 nm & 168.9 nm & Scanning\\

\end{tabular}

\label{datasetinfo}
\end{center}
\end{table}

\subsection{Results on Partially Synthetic Data} \label{Synthetic}
 
\begin{figure*}
\vspace{0.1in}
\begin{center}
  \includegraphics[width = 0.9in]{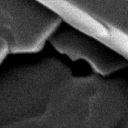}
  \includegraphics[width = 0.9in]{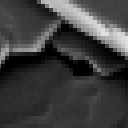}
  \includegraphics[width = 0.9in]{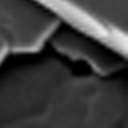}
  \includegraphics[width=0.9in]{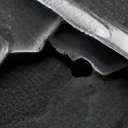}
  \includegraphics[width=0.9in]{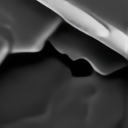}
  \includegraphics[width=0.9in]{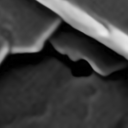}
  \includegraphics[width=0.9in]{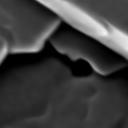}

\end{center}
\hspace{.45in} HR GT \hspace{.25in} Simulated LR \hspace{.25in} Bicubic \hspace{.3in} DPSRGAN \hspace{.35in} DPSR \hspace{.6in} PnP \hspace{.43in} \makered {\bf MDF-RAP} \makeblack

 \hspace{1.4in} 24.98 dB \hspace{.3in} 28.94 dB \hspace{.35in} 27.03 dB \hspace{.35in} 30.79 dB \hspace{.35in} 32.18 dB \hspace{.3in} \textbf{32.65 dB}

\caption{4x super-resolution reconstructions of a simulated LR EM image of pentacene crystals with $\mu = 0.8$ for MDF-RAP. Each shows a zoomed field of view 5.34$\mu$ wide. MDF-RAP performs well in terms of PSNR (and FRC in Figure~\ref{fig:FRC}) while DPSR produces edges that are even sharper than the edges in the HR GT.}
\label{fig:Pentacene}
\end{figure*}

In Figures~\ref{fig:8xNano}--\ref{fig:Pentacene}, we display a HR ground-truth image, the corresponding simulated LR data, the output of bicubic upsampling (as a baseline), DPSRGAN (when possible), DPSR, PnP (using a CNN trained on natural images as the prior agent), and MDF-RAP (using RAP and domain-specific training). 
Note that MDF-RAP is the only method that incorporates microscopy data in its training. None of Bicubic Interpolation, DPSRGAN, DPSR, and PnP have seen the microscopy data; we include these comparison points to illustrate the limits of domain-independent methods. The comparisons with MDF-RAP are meant to characterize improvements obtained by incorporating domain-specific data fusion in the prior model.

\begin{figure*}
\begin{center}
\begin{minipage}{3.7in}
  \includegraphics[width=3.7in]{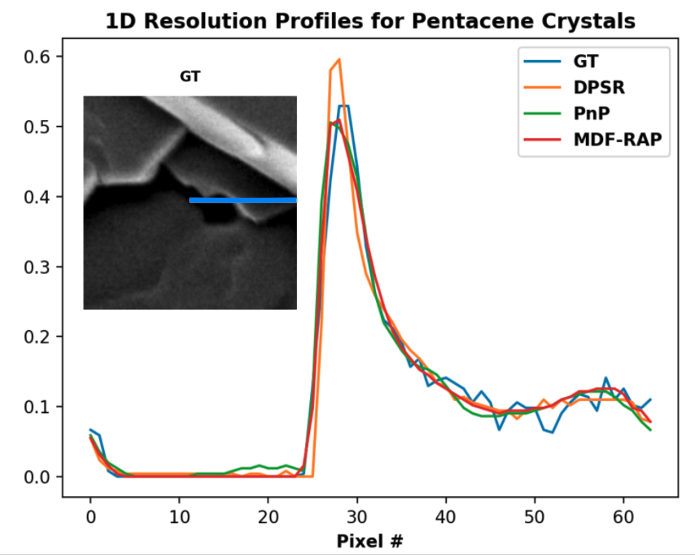}\\
     \mbox{\hspace{1.3in}} 1-D Resolution Profiles
\end{minipage}
  \hspace{0.2in}
\begin{minipage}{3in}
  \includegraphics[width = 1.2in]{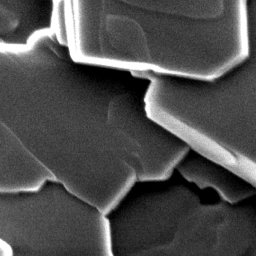}
  \includegraphics[width = 1.2in]{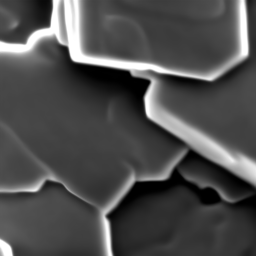}\\
  \mbox{\hspace{0.4in}} HR GT \hspace{0.7in} $\mu = 0.2$\\
  \mbox{\hspace{1.55in}} 31.48 dB\\[0.1in]
  \includegraphics[width = 1.2in]{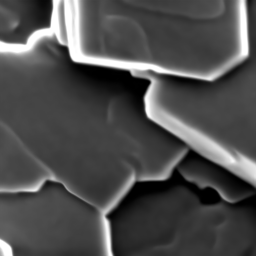}
  \includegraphics[width= 1.2in]{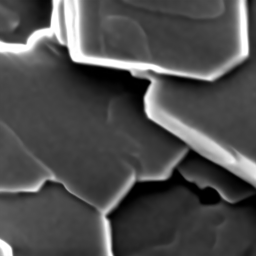}\\
  \mbox{\hspace{0.35in}} $\mu = 0.5$ \hspace{0.7in} $\mu = 0.8$\\
  \mbox{\hspace{0.35in}} 32.46 dB \hspace{0.6in}  \textbf{32.65 dB}
\end{minipage}

\end{center}

\caption{{\em Left:}  Plot of intensities along the blue segment for selected reconstructions from Fig~\ref{fig:Pentacene}.  DPSR produces the sharpest edge boundary, while PnP and MDR-RAP better approximate GT image on crystal edges.  {\em Right:}  Comparison of 4x super-resolution reconstructions of a simulated LR EM image of pentacene crystals at varying regularization levels. Note as $\mu$ increases, more detail is present in the final reconstruction. However this can introduce textural artifacts from the forward model into the final reconstruction.}
\label{fig:pentReg}
\end{figure*}

Figure~\ref{fig:8xNano} illustrates 8x super-resolution on gold nanorods, which are highly structured. Although 8x super-resolution is significantly underconstrained, MDF-RAP leverages the data-fidelity operator and a domain-specific CNN to produce a high-quality reconstruction.  Here the competing reconstructions each have significant shortcomings. Additionally note that no retraining was required to apply PnP or MDF-RAP at 8x, since changes to the resolution factor requires only a change of $L$ in the forward model rather than retraining the prior model.

In Figure~\ref{fig:4xecoli}, with 4x super resolution on data with less regularity and more fine detail, MDF-RAP produces the highest PSNR, although not by a significant margin.  However, MDF-RAP arguably provides the best visual compromise between clarity and fit to data among the competing methods. While DPSR produces lines and background structures with sharper edges than MDF-RAP, these sharp features do not always align with structures in the HR GT.

\begin{figure*}
\begin{center}
  \includegraphics[width=2.3in]{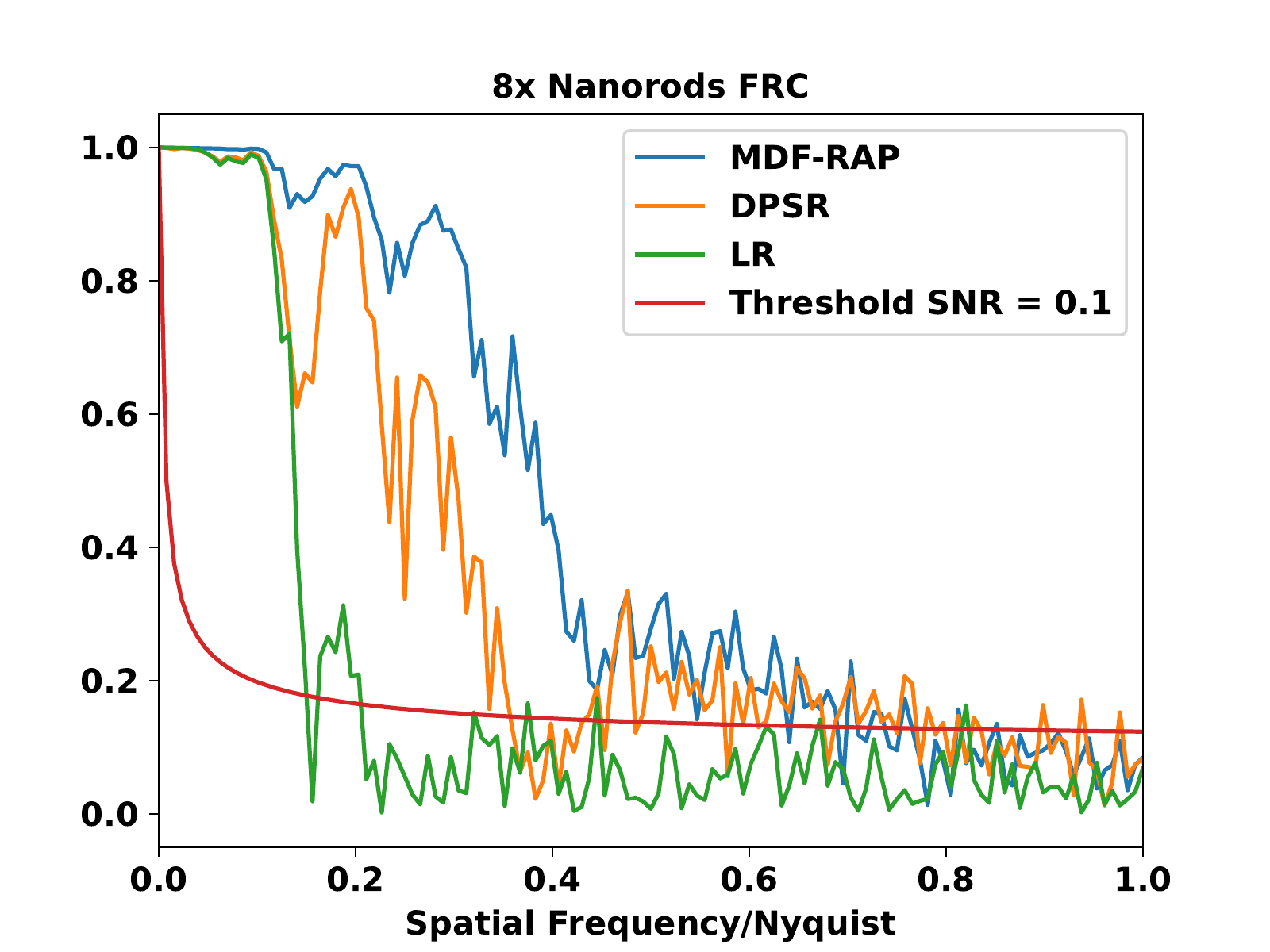}
  \includegraphics[width=2.3in]{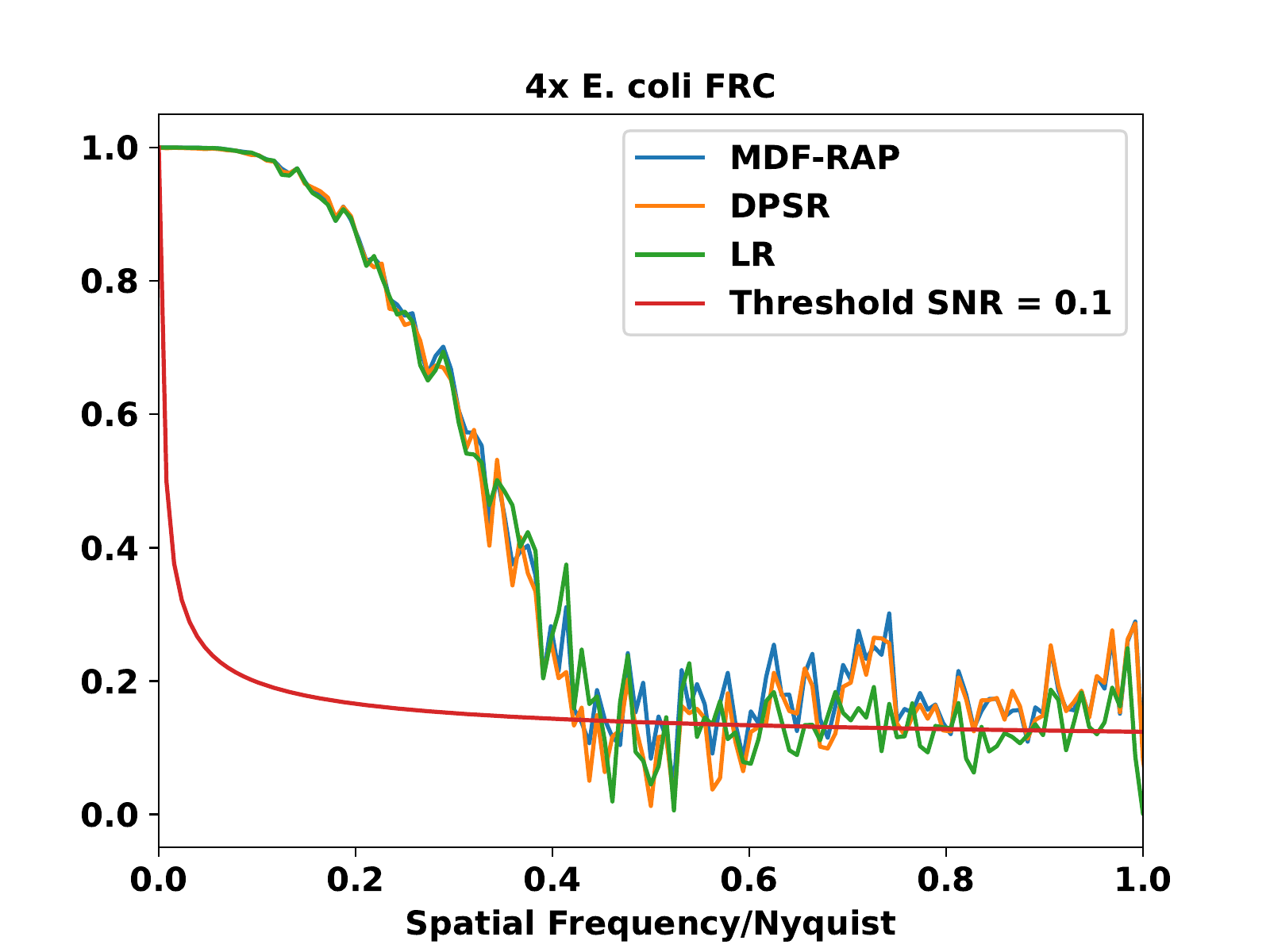}
  \includegraphics[width=2.3in]{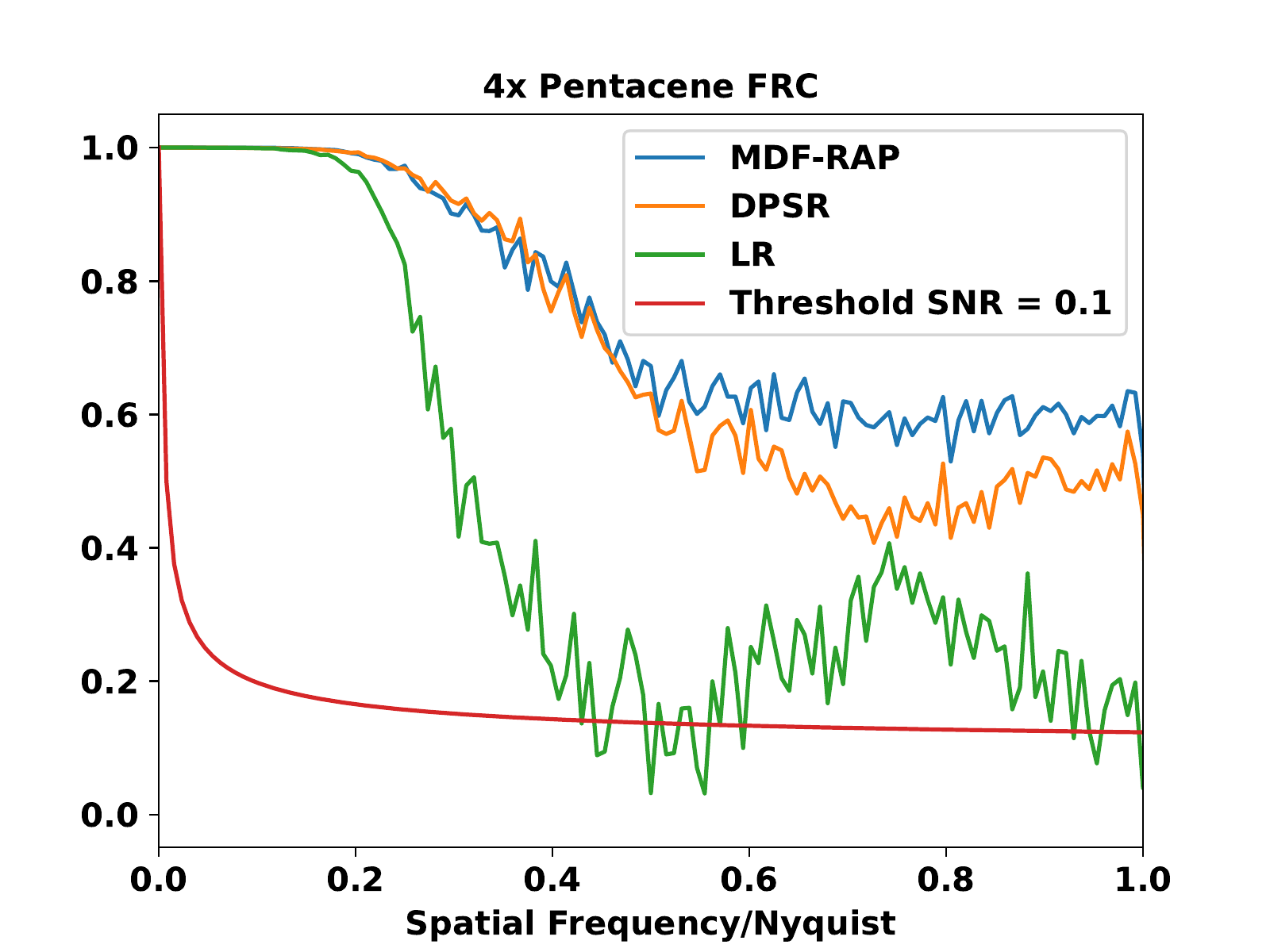}

\caption{ Fourier Ring Correlation plots for gold nanorods, {\em E. coli}, and pentacene.  FRC describes the correlation in frequency domain between two images; higher values indicate closer correlation.  The green lines are from the simulated LR images, with sharp decay near the sampling rates of $1/8$ for nanorods and $1/4$ for {\em E. coli} and pentacene. MDF-RAP outperforms DPSR on nanorods, but neither method improves significantly over the LR image for {\em E. coli}.  Consistent with the PSNR results in Figure 3, MDF-RAP outperforms DPSR on the pentacene example, despite the visual appeal of the DPSR result. \makeblack}
\label{fig:FRC}
\end{center}
\end{figure*}

In Figure~\ref{fig:Pentacene}, we show 4x super resolution reconstructions of pentacene crystals. MDF-RAP outperforms the other methods in terms of PSNR, with PnP a close second in this metric.  Visually, DPSR produces an image with edges that are sharper than in the HR GT image, which may account for the lower PSNR of DPSR relative to MDF-RAP. Such sharpness is clear in the 1D resolution profile shown in Figure~\ref{fig:pentReg}, as the peak of DPSR's plot is significantly higher than that of the GT.
To further investigate the the accuracy of MDF-RAP versus DPSR, we use Fourier Ring Correlation to compare these  reconstructions to the HR GT (discussion below and plots in Fig. \ref{fig:FRC}). 
\makeblack

In Table~\ref{tbl:PSNR-results}, we show average results for reconstructions of synthetic LR data at 4x interpolation.
For each dataset, we extracted 100 256x256 images and created corresponding simulated LR data.
These images were then passed into each interpolation method for reconstruction. 
Finally we collected the PSNRs relative to the original high-resolution image and averaged these across the 100 images to generate the values shown. 
We also include a comparison to using an image-tuned prior model with the forward model in \eqref{eq:5}, labeled as MDF.
As shown in Table~\ref{tbl:PSNR-results}, methods using an image-tuned prior outperform all other interpolation methods tested.

\makeblack

\begin{table}[h]
\caption{Average PSNRs over 100 images for \textit{E. coli}, Pentacene, and Nanorods datasets at 4x interpolation. On average, techniques incorporating MDF outperform all other methods.}
\begin{center}
\begin{tabular}{ c | c | c | c | c | c | c}
\multirow{2}*{Material} & \multirow{2}*{Bicubic} & DPSR- & \multirow{2}*{DPSR} & \multirow{2}*{PnP} & \multirow{2}*{MDF} & \makered {\bf MDF-} \makeblack \\
  &  & GAN &  &  &  & \makered {\bf RAP} \makeblack \\
 \hline
  Nanorods & 32.43 & 29.87 & 34.55 & 34.21 & \textbf{34.97} & 34.95 \\
   \textit{E. coli} & 19.98 & 14.27 & 19.69 & 19.95 & \textbf{20.16} & 20.05 \\ 
 Pentacene & 30.00 & 28.84 & 32.65 & 33.78 & 34.09 & \textbf{34.29} 
\end{tabular}
\label{tbl:PSNR-results}
\end{center}
\end{table}

 We investigate the effect of the prior model on image reconstruction in two ways.  
First, the weight $\mu$ in the MACE formulation can be used to tune the regularization strength of the prior.  As $\mu$ increases, the forward model is weighted more and thus the amount of regularization is reduced.  In Figure~\ref{fig:pentReg}, we compare reconstructions as $\mu$ varies and note the increased smoothing that occurs as $\mu$ is decreased.  We used $\mu = 0.8$ for the  reconstructions in Figures~\ref{fig:8xNano}--\ref{fig:Pentacene} to  preserve image detail in the final reconstructions.

\begin{figure*}
\begin{center}
  \includegraphics[width =  2.3in]{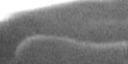}
  \includegraphics[width = 2.3in]{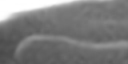}
  \includegraphics[width = 2.3in]{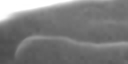}
\end{center}
   \hspace{1.0in} HR GT 
   \hspace{1.25in} PnP (no RAP) 35.06 dB
   \hspace{0.9in} MDF w/o RAP 35.96 dB

\begin{center}
    \includegraphics[width=2.2in]{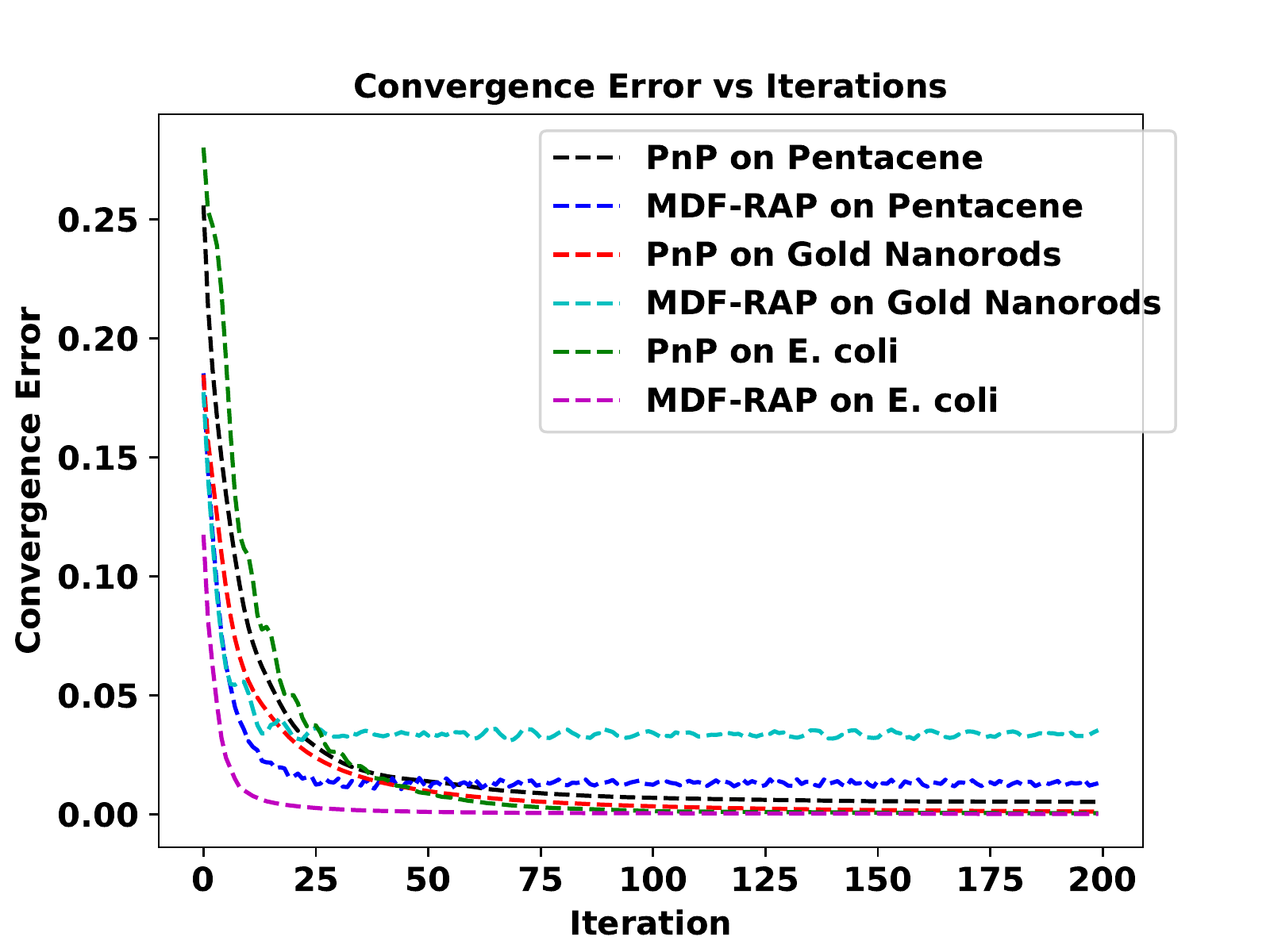}\mbox{\hspace{0.1in}}
  \includegraphics[width = 2.3in]{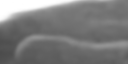}
  \includegraphics[width = 2.3in]{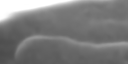}
\end{center}
 \hspace{0.7in} Convergence Error
 \hspace{0.95in} PnP plus RAP 35.27 dB
 \hspace{0.9in} \makered {\bf MDF-RAP} {\bf 36.13 dB}

\caption{ {\em Top row and bottom right: } 4x super-resolution reconstructions of a simulated LR image of pentacene crystals as in Figures~\ref{fig:Pentacene} and \ref{fig:pentReg}.  Here PnP (with or without RAP) uses a DnCNN denoiser trained only on natural images, while MDF  (with or without RAP) uses a domain-specific denoiser.  PnP (no RAP) has a herringbone texture, while MDF w/o RAP has 4x4 blocky artifacts, neither of which appears in the HR GT or the reconstructions with RAP.   MDF with or without RAP produces a clearer edge than the PnP reconstructions.  
{\em Bottom left:} Convergence Error (equation~\eqref{eq:conv-error}) plots for 4x PnP and 4x MDF-RAP on pentacene, gold nanorods, and \textit{E.~coli} datasets. All methods converge with under 5\% error, with fastest convergence on the \textit{E. coli} dataset. \makeblack} 
\label{fig:pentRAP}
\end{figure*}

Second, we investigate domain-specific training and RAP one at a time in Figure~\ref{fig:pentRAP}. While the use of RAP is explicitly a change in the forward agent, it is equivalent to a change in the prior.  The top 2 reconstructions show PnP and MDF without RAP, both of which yield unnatural artifacts in texture.  PnP has a herringbone pattern, while MDF without RAP has 4x4 blocks that do not match the surrounding background.  The results with RAP in the bottom two images do not have these artifacts.  The MDF-RAP reconstruction provides the best clarity and PSNR.    

A possible explanation for the artifacts in the reconstructions without RAP is that the equilibrium solution in equation \eqref{eq:2} requires an update from the forward model that is counterbalanced by an equal but opposite update from the prior model.  In the case of the standard backprojector, the update in \eqref{eq:5} has the form of adding a blocky image in the range of $A^T$ to the existing reconstruction.  This must be balanced by an update from the neural network that results in the same output image.  A blocky update from a neural network is unlikely unless the underlying image itself has some structure that promotes such an update.  We hypothesize that the artifacts seen in the no-RAP reconstructions arise from this need to balance the forward and prior updates and that the use of the smoother RAP update reduces the need for such artificial structures in the final reconstruction.  
\makeblack

In the bottom left of Figure \ref{fig:pentRAP}, we plot the average convergence error in \eqref{eq:conv-error} as a function of iteration over these 100 256x256 images.  For each dataset, MDF-RAP converged with under 5\% error. 
The gold nanorods dataset at 8x superresolution leads to the highest convergence error, which is likely due to the existence of multiple data-consistent reconstructions at this level of superresolution.

In Figure~\ref{fig:FRC}, we use Fourier Ring Correlation (FRC) \cite{BanterleFRC} to estimate image resolution. FRC measures the correlation between the frequency spectra of 2 images, with the frequency domain subdivided into concentric rings to define spectral correlation as a function of spatial frequency. The effective image resolution is determined by the intersection of the FRC curve with a given threshold curve, with higher FRC values indicating closer match to the target image. We use code from \cite{frc_code} to generate the plots in Figure~\ref{fig:FRC}. 

The plots in Figure~\ref{fig:FRC} show that MDF-RAP recovers significantly more of the high frequency components of the 8x gold nanorods image than DPSR, which in turn captures more than the LR image.  This is consistent with the visual results of Figure~\ref{fig:8xNano}.   MDF-RAP, DPSR, and the LR image all have very similar FRC curves for the \textit{E. coli} images, again consistent with Figure~\ref{fig:4xecoli}.  In the case of pentacene, the FRC curve for MDF-RAP stays noticeably higher than DPSR in the upper frequency range. This is consistent with the PSNR in Figure~\ref{fig:Pentacene} despite the visual appeal of the DPSR result.  The intersection of the FRC curves with the threshold SNR curve indicates that DPSR achieves an effective super-resolution of roughly 2x on gold nanorods, while MDF-RAP achieves a super-resolution in the range of 3x to 5x.  Neither method provides a significant improvement on the \textit{E. coli} image, and both achieve something close to the target 4x super-resolution on the pentacene image. 

\makeblack

\subsection{Results on Measured Data} \label{Real Data}

\begin{figure*}
    \begin{center}
    \includegraphics[width = 0.9in]{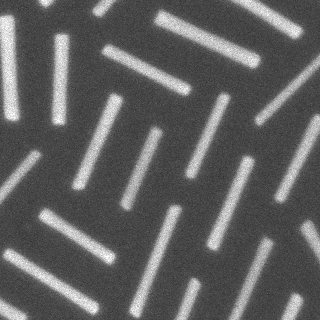}
    \includegraphics[width = 0.9in]{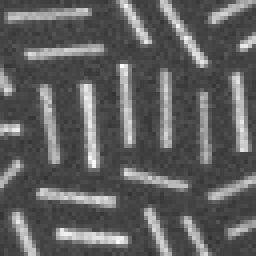}
    \includegraphics[width = 0.9in]{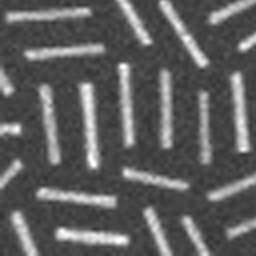}
    \includegraphics[width = 0.9in]{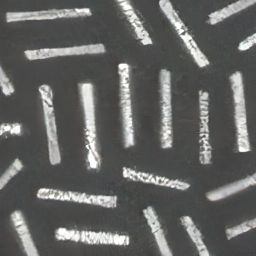}
    \includegraphics[width = 0.9in]{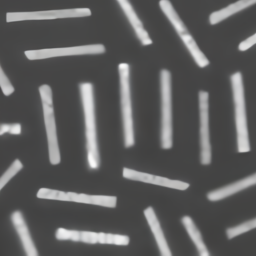}
    \includegraphics[width = 0.9in]{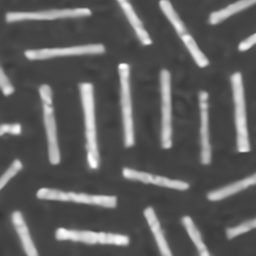}
    \includegraphics[width = 0.9in]{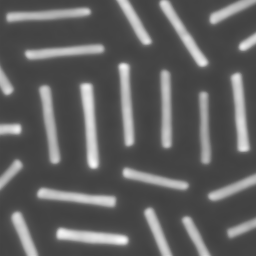}
    \end{center}

\hspace{.3in} HR Example \hspace{.1in} Measured LR \hspace{.25in} Bicubic \hspace{.3in} DPSRGAN \hspace{.35in} DPSR \hspace{.6in} PnP \hspace{.4in} \makered MDF-RAP \makeblack
    
    \caption{4x super-resolution reconstructions of a measured LR EM image of gold nanorods.  In this case there is no aligned HR image; the left panel is for qualitative comparison.  Each panel shows a field of view 352 nm wide, and MDF-RAP uses $\mu = 0.2$. While all methods reconstruct the nanorods' shape, only MDF-RAP reconstructs the nanorods without internal gaps and maintains the slightly curved ends characteristic of gold nanorods.}
    \label{fig:real4xNano}
\end{figure*}

\begin{figure*}
\vspace{0.1in}
    \begin{center}
    \includegraphics[width = 0.9in]{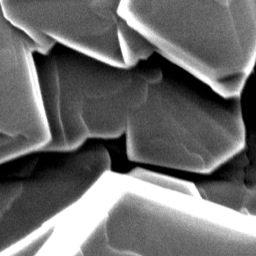}
    \includegraphics[width = 0.9in]{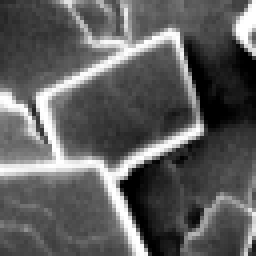}
    \includegraphics[width = 0.9in]{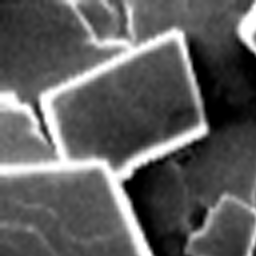}
    \includegraphics[width = 0.9in]{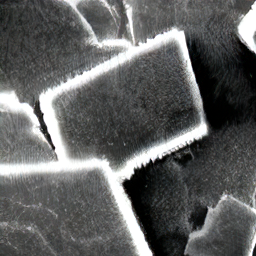}
    \includegraphics[width = 0.9in]{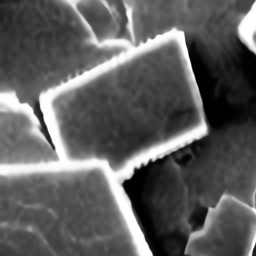}
    \includegraphics[width = 0.9in]{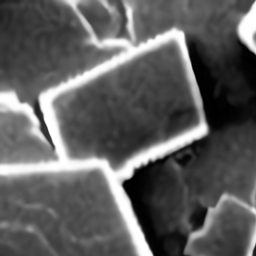}
    \includegraphics[width = 0.9in]{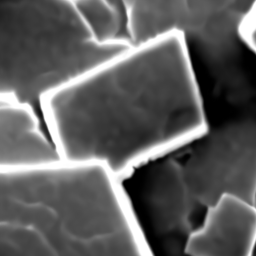}
    \end{center}

\hspace{.3in} HR Example \hspace{.1in} Measured LR \hspace{.25in} Bicubic \hspace{.3in} DPSRGAN \hspace{.35in} DPSR \hspace{.6in} PnP \hspace{.4in} \makered MDF-RAP \makeblack
    
    \caption{4x super-resolution reconstructions of a measured LR EM image of pentacene crystals. In this case there is no aligned HR image; the left panel is for qualitative comparison.  Each panel shows a field of view 10.67$\mu$ wide, and MDF-RAP uses $\mu = 0.2$. Note the significant aliasing artifacts along the crystal edges in all non-MDF reconstructions. }
    \label{fig:realpent}
\end{figure*}

In Figures \ref{fig:real4xNano}--\ref{fig:realpent}, we show results using measured LR data as input.  In this case there is no paired HR data for quantitative comparison, so we provide a measured HR image of a similar specimen for visual comparison.  

In Figure \ref{fig:real4xNano}, with 4x superresolution of gold nanorod images, the data is not severely undersampled, so each method is able to reconstruct the shape of the nanorods. However, relative to the other methods, MDF-RAP provides a more faithful reconstruction of the nanorod interiors and ends while removing background noise.

In Figure \ref{fig:realpent}, the majority of the methods produce aliasing artifacts along the edge of the crystal. MDF-RAP minimizes these artifacts relative to the other methods and again provides a good balance between clarity and realistic texture.

 \subsection{Speed-Up and Computational Complexity}
\label{Complexity}
\makeblack

In Table~\ref{tbl:speedup-results}, we examine the speed-up in acquisition time by using MDF-RAP. The speed-up is calculated by taking the ratio of the pixels necessary for a HR FoV to the sum of the pixels in the LR FoV and the pixels in the HR training data. 
\begin{align*}
    \text{Speed-Up} = \frac{\text{HR Reconstruction Pixels}} {\text{Acquired LR pixels + HR Training Pixels}}
\end{align*}
In the ideal case, in which a domain-specific CNN denoiser is already trained, the acquisition speed-up for $L$x interpolation is $L^2$.  In the cases shown in Table~\ref{tbl:speedup-results} we include the HR data acquisition needed for CNN training, so the actual speed-up ranges from roughly 50\% to 62\% of the ideal.

A limitation of our method is reconstruction time. Running 20 iterations of MDF-RAP on a 256 x 256 GT image takes 32.6 seconds on a CPU, while DPSR takes 12.2 seconds. This ratio is preserved on a GPU, where MDF-RAP takes 18.8 seconds and DPSR takes 7 seconds. The longer running time of MDF-RAP is likely caused by repeated application of the prior model. One could accelerate the MDF-RAP code by incorporating parallelization and better managing memory, which we leave as future work.  In contrast, an advantage of MDF-RAP over DPSR is that the CNN for MDF-RAP is trained on high-resolution images, so does not require retraining for new super-resolution levels or point spread functions.
\makeblack

\begin{table*}[ht]
\caption{Data acquisition speedup with MDF-RAP. The speedup factor is calculated by taking the ratio of the number of HR reconstructed pixels to the sum of the acquired LR pixels and HR training pixels}
\begin{center}
\vspace{0.1in}
\normalsize
\begin{tabular}{ c | c | c | c | c | c }
 Material & Interpolation Factor & LR data acquired & HR training data & Reconstructed HR data & Speed-Up \\
 \hline
 Nanorods & 5x & 2048 x 1388 & 1232 x 1367 & 10240 x 6940 &  15.70x\\    
 Nanorods & 8x & 2048 x 1388 & 1232 x 1367 & 16384 x 11104 &  40.19x\\ 
  \textit{E. coli} & 4x & 7404 x 7666 & 5049 x 9827 & 29616 x 30664 &  8.54x \\ 
 Pentacene & 4x & 1280 x 755 & 1280 x 755 & 5120 x 3020 & 10.05x\\  

\end{tabular}

\label{tbl:speedup-results}
\end{center}
\end{table*}

\section{Conclusion} \label{Conclusion}
 
We introduced a Multi-Resolution Data Fusion framework with a Relaxed Adjoint Projection and a domain-specific neural network prior operator. RAP leads to reduced textural artifacts, is easy to implement, and is shown to be equivalent to using the standard data fitting operator with a modified prior.  The domain-specific neural network prior operator is trained on a limited set of high-resolution images that do not require pairing with low-resolution images.  

In a set of experiments, MDF-RAP improves quality relative to existing methods while maintaining data fidelity, accurately resolving sub-pixel-scale features, and providing speed-ups of 8 to 20 times relative to imaging a full high-resolution image. 
MDF-RAP can be used at multiple super-resolution factors without additional training, and by changing the forward model, it can be used for multiple image acquisition models.  This modularity is an important  strength in that each component can be used for multiple applications.

More philosophically, the goal of MDF-RAP (and of regularized inversion generally) is to bias the reconstruction to compensate for a set of imperfect measurements.  The comparisons between the MDF-RAP and PnP reconstructions show that the results are improved by using domain-specific training data.  This bias-variance tradeoff can be pushed too far in the case of many reasonable reconstructions that are consistent with measured data, so any use of super-resolution methods requires thoughtful use and some form of validation.   Even so, speed-ups of 8 to 20 times relative to imaging a full high-resolution image justify further investigation and use.  
\makeblack

\appendix

 In this appendix, we prove Theorem 1, which shows that the equilibrium solutions of RAP are identical to those using a standard backprojector with an alternative averaging operator. We then provide a series of lemmas that culminate in Theorem 2, which shows that the use of RAP with a given denoiser is equivalent to using the standard backprojector and an alternative prior. Finally, we prove Theorem 3, which shows that Algorithm 1 with RAP converges to a fixed point.
\makeblack 

We begin with a proposition necessary to define several maps. 

\medskip
\begin{propn} \label{propn:boyd}
Let $\phi$ be maximally monotone. Then $(I+\phi)^{-1}$ is globally defined, single-valued, and nonexpansive. 
\end{propn}

\vspace{0.1in}
\begin{proof}
See \cite[section 6]{boyd}. 
\end{proof}

\vspace{0.1in}
In the theorem below, we use maps $\textbf{F}_i$ that are implicitly defined in the sense that the map $\phi_i$ is evaluated at the output of the corresponding map $\textbf{F}_i$.  This is a generalization of the condition satisfied by a proximal map and is equivalently written as $(I+\phi_i)^{-1}$, which is known as the resolvent of $\phi_i$, as in Theorem~\ref{thm:FG-equiv}.  

\vspace{0.1in}
\begin{thm} \label{thm:FG-equiv}
Assume that each of $\phi_i$ and $R_i\phi_i$ are maximal monotone functions from ${\mathbb R}^n$ to itself for each $i=1, \ldots, K$, where each $R_i$ is an $n \times n$ matrix with $\sum_i R_i$ invertible.  Define

\begin{itemize}
    \item $\textbf{F}_i(v_i) = v_i - \phi_i ( \textbf{F}_i(v_i) ) = (I + \phi_i)^{-1}(v_i)$
    \item $\textbf{F}^R_i(v_i) = v_i - R_i \phi_i(\textbf{F}^R_i(v_i)) = (I + R_i \phi_i)^{-1}(v_i)$
    \item $\textbf{G}_i(\textbf{v}) = \frac{1} {K} \sum_i v_i$
    \item $\textbf{G}^R_i(\textbf{v}) = (\sum_i R_i)^{-1} (\sum_i R_i v_i)$
\end{itemize}
Then there is a map from solutions  $\textbf{v}^*$ of 
$$\textbf{F}^R(\textbf{v}^*) = \textbf{G}(\textbf{v}^*)$$
to solutions $\textbf{\^ v}^*$ of 
$$\textbf{F}(\textbf{\^ v}^*) = \textbf{G}^R(\textbf{\^ v}^*)$$
such that for each such pair, $\textbf{G}(\textbf{v}^*) = \textbf{G}^R(\textbf{\^ v}^*)$.  There is also such a map from $\textbf{\^ v}^*$ to $\textbf{v}^*$.  Moreover, the common value $x^*$ in the stacked vector $\textbf{G}(\textbf{v}^*)$ satisfies $\sum_i R_i \nabla f_i(x^*) = 0$.
\end{thm}

\vspace{0.1in}
Note that $\textbf{G}(\textbf{v}^*)$ is formed by stacking copies of the consensus solution $x^*$, so this theorem says that the two formulations in (i) and (ii) have exactly the same set of consensus solutions.  

\vspace{0.1in}
\begin{proof}
Assume that $\textbf{F}^R(\textbf{v}^*) = \textbf{G}(\textbf{v}^*)$, and define $x^*$ to be the identical entries of $\textbf{G}(\textbf{v}^*)$, so that $F_i(v_i^*) = x^*$ for each $i$. Applying $\textbf{G}$ to both sides yields $\textbf{G}(\textbf{F}^R(\textbf{v}^*)) = \textbf{G}^2(\textbf{v}^*) = \textbf{G}(\textbf{v}^*)$. Expanding using the definitions of $\textbf{F}^R_i$ and $\textbf{G}_i$ yields
\begin{align}
    \frac{1} {K} \sum_i (v_i^* - R_i \phi_i(\textbf{F}_i(v_i^*))) = \frac{1} {K} \sum_i v_i^*.
\end{align}
Multiplying by $K$ and canceling the sum of the $v_i^*$ gives 
\begin{align} \label{eq:R-phi}
\sum_i R_i \phi_i(x^*) = 0.
\end{align}
Conversely given $x^*$ such that $\sum_i R_i \phi_i(x^*) = 0$, define $v_i^* = x^* +R_i\phi_i(x^*)$ for all $i$. 
We will show that $\textbf{F}^R(\textbf{v}^*) = \textbf{G}(\textbf{v}^*)$.  Since $\textbf{F}^R_i(v_i) = (I+R_i\phi_i)^{-1}(v_i)$, we have
\begin{align}
    \textbf{F}^R_i(v_i^*) = (I+R_i \phi_i)^{-1} (x^* + R_i \phi(x^*)) = x^*.
\end{align}
Also,
\begin{align}
    \textbf{G}(\textbf{v}^*)_i &= \frac{1} {N} \sum_i (x^* + R_i \phi (x^*))  \\
    &= x^* + \frac{1} {N} \sum_i  R_i\phi_i(x^*).
\end{align}
Note that the second term in this sum is 0 by assumption, so $\textbf{G}(\textbf{v}^*)_i = x^*$ for all $i$ and hence 
$\textbf{F}^R(\textbf{v}^*) = \textbf{G}(\textbf{v}^*)$.

Assume now that $\textbf{F}(\hat{\textbf{v}}^*) = \textbf{G}^R(\hat{\textbf{v}}^*)$, and let $\hat x^*$ be the identical entries of $\textbf{F}(\hat{\textbf{v}}^*)$. As before, $\textbf{G}^R(\textbf{F}(\hat{\textbf{v}}^*)) = \textbf{G}^R(\hat{\textbf{v}}^*)$ and this with the definitions yields
\begin{align}
    \left(\sum R_i\right)^{-1}  \sum R_i(\hat{v}_i^*  &  - \phi_i(\textbf{F}_i(\hat{v}_i^*)))  \nonumber\\
    &= \left(\sum R_i\right)^{-1}  \left(\sum R_i \hat{v}_i^*\right).
\end{align}
Applying $(\sum R_i)$ to both sides, canceling $\sum R_i \hat{v}_i^*$, and using $\textbf{F}_i(\hat{v}_i^*) = \hat x^*$ gives $\sum_i R_i \phi_i(\hat x^*) = 0$.

Conversely given $\hat x^*$ such that $\sum R_i \phi_i( x^*) = 0$, define $\hat{v}_i^* = \hat x^*+\phi_i(x^*)$ for all $i$. A calculation similar to the previous case shows that  $\textbf{F}(\hat{\textbf{v}}^*) = \textbf{G}^R(\hat{\textbf{v}}^*)$.

Hence for each $\textbf{v}^*$ satisfying (i), the identical entries $x^*$ of $\textbf{G}(\textbf{v}^*)$ satisfy \eqref{eq:R-phi}.  
This condition then determines $\hat{\textbf{v}}^*$ satisfying (ii) and so that $x^*$ is the common entry in $\textbf{G}^R(\textbf{v}^*)$.  
The map from $\hat{\textbf{v}}^*$ to $\textbf{v}^*$ is the same in reverse.
\end{proof}

\vspace{0.1in}

\begin{lemma} \label{eq:lemmaFform}
The map $F$ in \eqref{eq:5} can be expressed as either 
\begin{enumerate}
    \item $F(x) = (I + \nabla f)^{-1}(x)$
    \item $F(x) = (I - r \nabla f)(x)$,
\end{enumerate}
where $\sigma^2 = \frac{\sigma_\lambda^2} {\sigma_w^2}$, $f = \frac{\sigma^2} {2} \|y-Ax\|_2^2$ and $r = 1/(1+\sigma^2 L^2)$.  
\end{lemma}

\medskip
\begin{proof}
We'll first establish the form in 1. Note that the first order optimality condition for $F(x) = \argmin_v \left\{ f(v) + \frac{1}{2} \|v - x\|^2 \right\}$ is $\nabla f(v) + v-x = 0$.
Solving for $v$ gives $v^* = (I+\nabla f)^{-1}(x)$. Since $f$ is a positive, semi-definite quadratic penalty, its subdifferential is maximal monotone, hence this inverse is well-defined by Proposition 1. 

Using $\nabla f(v) = \sigma^2 A^T (Av - y)$ in the first order optimality condition above and isolating $v^*$ gives
\begin{align} \label{eq:backsub}
     v^* &= x - \sigma^2 A^T(Av^*-y).
\end{align}
Therefore, $v^*$ is of the form $v^* = x+A^Tz$ for some $z$. Using this form of $v^*$ in \eqref{eq:backsub} gives
\begin{align}
x+A^Tz &= x - \sigma^2 A^T (A (x+A^Tz) - y).
\end{align}
Some algebra and $A A^T = L^2 I$ gives 
\begin{align}
     A^T(y-Ax) &= (1+L^2\sigma^2)A^Tz,
\end{align}
with a solution of $z = \frac{\sigma^2} {1+\sigma^2L^2} (y-Ax)$. We substitute this into $v^* = x+A^Tz$ to obtain
$F(x) = v^* = x + r \sigma^2 A^T(y-Ax)$, or $(I-r\nabla f)(x)$, where $r = \frac{1} {1+\sigma^2L^2}$. 
\end{proof}

\vspace{0.1in}
\vspace{0.1in}

\begin{lemma} \label{eq:liplemma}
Suppose Lip$(\psi) \leq \alpha < 1$.  Then $\Psi = I + \psi$ is invertible and
\begin{align}
    \text{Lip}(\Psi^{-1}) & \leq \frac{1}{1-\alpha}\\
    \text{Lip}(I - \Psi^{-1}) & \leq \frac{\alpha}{1-\alpha}
\end{align}
\end{lemma}

\medskip
\begin{proof}
If $\psi$ is Lipschitz with constant $\alpha$, then $\Psi = I + \psi$ is also Lipschitz.
Consequently $\Psi$ will be differentiable almost everywhere by Rademacher's theorem.
The forward and reverse triangle inequalities imply
\begin{align}
    (1-\alpha)\|x-z\| \leq \|\Psi(x)-\Psi(z)\| \leq (1+\alpha)\|x-z\|.
\end{align}
These bounds show that the singular values of $\Psi$ are bounded away from 0, which implies its Jacobian is of full rank. The Lipschitz Inverse Function Theorem \cite{clarke} implies that $\Psi$ is invertible. Defining $w = \Psi(x)$ and $v = \Psi(z)$ transforms the bounds on $\|\Psi(x)-\Psi(z)\|$ to
\begin{align}
    (1-\alpha)\|\Psi^{-1}(w)&-\Psi^{-1}(v)\| \leq \|w-v\| \nonumber\\
    &\leq (1+\alpha)\|\Psi^{-1}(w)-\Psi^{-1}(v)\|.
\end{align}
Dividing the left hand side of the inequality by $1-\alpha$ yields Lip$(\Psi^{-1}) \leq 1/(1-\alpha)$.

Now fix $v_1$ and $v_2$, and let $w_j = \Psi(v_j) = v_j + \psi(v_j)$, so that $(I-\Psi^{-1})(w_j) = \psi(v_j)$. Using the Lipschitz constants for $\psi$ and $\Psi^{-1}$ gives
\begin{align}
\|(I-\Psi^{-1})(w_1)&-(I-\Psi^{-1})(w_2)\| \\
    &= \|\psi(v_1) - \psi(v_2)\| \\
    &\leq \alpha \|v_1-v_2\| \\
    &= \alpha \|\Psi^{-1}(w_1)-\Psi^{-1}(w_2)\| \\
    &\leq \frac{\alpha} {1-\alpha}\|w_1-w_2\|.
\end{align}
\end{proof}
\vspace{0.1in}

\begin{lemma} \label{eq:lemmaFRform}
\noindent
Assume $\sigma^2 < 1/L^2$ and let $f$ and $r$ be as in Lemma~\ref{eq:lemmaFform} with this $\sigma^2$. Then there exist constants $\delta, C_1 > 0$ such that if $R$ is a matrix satisfying $\|R-I\| < \delta$, then there exists a matrix $\tilde R$ depending on $R$ so that 
\begin{itemize}
    \item  $\|\tilde R - I \| \leq C_1 \|R - I\|$
    \item $\tilde R \nabla f$ is maximal monotone
\end{itemize}
and so that 
\begin{align} \label{eq:R-equiv}
x - rR \nabla f(x) = (I + \tilde R \nabla f)^{-1}(x). 
\end{align}
\end{lemma}

\medskip
\begin{proof}
Define $W = \sigma^2 A^TA$.  Note that $A^T A$ can be factored as a projection followed by scaling by $L^2$, hence $\|rW\| = \sigma^2 L^2 / (1 + \sigma^2 L^2) < 1 / (1 + \sigma^2 L^2)$ by assumption. 
Thus there exists $\delta_1 > 0$ so that if $\|R-I\| < \delta_1$, then $\|rWR\| < 1$, hence $(I - rWR)$ is invertible by Lemma~\ref{eq:liplemma}.  As motivated below, let
\begin{align} \label{eq:Rtilde}
    \tilde{R}x = \begin{cases} rR(I-rWR)^{-1} x & \text{ for } x \in \text{range}(A^T)\\
    Rx & \text{ for } x \in \text{null}(A).
    \end{cases}
\end{align}
Since the orthogonal complement of $\text{range}(A^T)$ is $\text{null}(A)$, we can extend by linearity to all of $\mathbb{R}^n$. 

Since $A A^T = L^2 I$, induction shows that $W^k A^T = (\sigma^2 L^2)^k A^T$. Since $\sigma^2 L^2 < 1$, we can expand the first part of \eqref{eq:Rtilde} at $R=I$ using a convergent power series to get
\begin{align}
    r(I - rW)^{-1} A^T &= r \sum_{k=0}^\infty r^k W^k A^T\\
    &= r \sum_{k=0}^\infty (r\sigma^2 L^2)^k A^T\\
    &= \frac{r}{1 - r\sigma^2 L^2} A^T.
\end{align}
Since $r = 1/(1+\sigma^2 L^2)$, this is $A^T$.

The same idea shows that for $R$ sufficiently close to the identity, say $\|R - I\| < \delta_2$ for some $\delta_2 \in (0,1)$, $\tilde R = \tilde R(R)$ restricted to $\text{range}(A^T)$ can be written as a power series in $R$ and satisfies $\tilde R(I) = I$. Expanding this power series about $R=I$ gives constants $c_1, c_2 > 0$ so that 
\begin{align}
    \| \tilde{R} - I \| \leq  (c_1 + c_2\|R-I\|)\|R-I\|.
\end{align}
Define $C_1 = \max(1, c_1 + c_2)$. Since $\|R-I\| \leq 1$, we have
\begin{align} \label{eq:RIC}
    ||\tilde{R}-I|| \leq  C_1 ||R-I||
\end{align}
on $\text{range}(A^T)$, hence on all of $\mathbb{R}^n$ since $\tilde R = R$ on $\text{null}(A)$.

We now show that $ \tilde{R}\nabla f$ is maximal monotone. Note that 
\begin{align}
   \tilde{R}\nabla f(x) - \tilde{R} \nabla f(w) = \sigma^2 \tilde{R}A^TA(x-w),
\end{align}
so $\tilde{R}\nabla f$ is maximal monotone when $\tilde{R}A^TA$ is positive semi-definite.
Note that if $x \in \text{null}(A)$, then $x^T\tilde{R}A^TAx =0$. If $x \in \text{range}(A^T)$, then $x = A^Tz$ for some $z \in \mathbb{R}^n$. Substituting this for the right-side $x$ gives
\begin{align}
    x^T\tilde{R}A^TAx = x^T \tilde{R} A^T A A^T z.
\end{align}
Since $AA^T = L^2I$, this is $L^2x^T\tilde{R}x$. By reducing $\delta_2$ if needed, \eqref{eq:RIC} implies that $\tilde R$ is positive definite. Extending by linearity shows that $\tilde{R} A^T A$ is positive semi-definite, so $\tilde{R}\nabla f$ is maximal monotone. Let $\delta = \min\{\delta_1, \delta_2\}$. 

Finally, let $\tilde F(x) = x-rR \nabla f(x)$, so $\tilde{F}(x) = (I-rR \nabla f)(x)$. Then $\tilde{F}(x)=(I+\tilde{R}\nabla f)^{-1}(x)$ exactly when 
\begin{align}
    (I + \tilde{R} \nabla f) \circ (I-rR\nabla f)(x) = x
\end{align}
Expanding and rearranging gives
\begin{align} \label{eq:tildeR1}
    \tilde R \nabla f \circ (I - rR \nabla f) = rR \nabla f.
\end{align}
Let $p = \sigma^2A^Ty$, so that $\nabla f(x) = Wx - p$. 
Using this in \eqref{eq:tildeR1} gives 
\begin{align}
    \tilde R(Wx - rWR(Wx-p) -p) = rR(Wx-p),
\end{align}
and collecting terms gives
\begin{align} \label{eq:tildeR2} 
    \tilde R (I - rWR)(Wx - p) = rR(Wx - p).
\end{align}
Since $Wx-p$ maps $\mathbb{R}^n$ onto $\text{range}(A^T)$, this is equivalent to $\tilde R (I - rWR) = rR$ on $\text{range}(A^T)$, which is consistent with \eqref{eq:Rtilde}.  Hence $\tilde{R}$ as defined in \eqref{eq:Rtilde} satisfies \eqref{eq:R-equiv}, thus completing the proof.
\end{proof}

\vspace{0.1in}

\begin{lemma} \label{eq:lemmaphiRform}
Let $f$ be as in Lemma \ref{eq:lemmaFform} and let $\phi$ be a Lipschitz and strongly maximal monotone function with Lipschitz constant $k$. Then there exists a constant $\delta > 0$ such that if $R$ is a matrix satisfying $\|R-I\| < \delta$, then $\tilde R^{-1} \phi$ is maximal monotone, where the matrix $\tilde R$ is from Lemma~\ref{eq:lemmaFRform}.  Moreover, there exists a function $\Phi_R$ and constant $C$ such that 
$$\text{Lip}(\Phi_R-I) \leq C \|R - I\|$$
and so that 
$$ (I+ \phi)^{-1} \circ \Phi_R = (I + \tilde R^{-1} \phi)^{-1}.$$
\end{lemma}

\medskip
\begin{proof}
It suffices to show $\Phi_R(x) = (I+\phi)(I+\tilde{R}^{-1}\phi)^{-1}(x)$ has the desired properties. We add and subtract $\phi$ and factor out $(I+\phi)^{-1}$ to obtain 
\begin{align}
    \Phi_R &= (I+\phi)(I+\phi + (\tilde{R}^{-1}-I) \phi)^{-1}\\
    &= (I+\phi)[(I + (\tilde{R}^{-1}-I) \phi (I + \phi)^{-1})(I+\phi)]^{-1}\\
    &= (I+(\tilde{R}^{-1}-I)\phi(I+\phi)^{-1})^{-1}. 
\end{align}
Hence $\Phi_R = (I+\psi)^{-1}$ with $\psi = (\tilde{R}^{-1}-I) \phi (I + \phi)^{-1}$. By Lemma~\ref{eq:lemmaFRform},
$\|\tilde{R} - I\| \leq C_1 \|R-I\|$. Restrict $R$ so that this is less than 1/2, so by Lemma~\ref{eq:liplemma} with $\Psi = \tilde R$, 
\begin{align}  
    \|\tilde{R}^{-1}-I\| \leq 2 C_1 \|R-I\| \label{eq:I-tildeR}.
\end{align}  
Let $d_1 = Lip((I+\phi)^{-1})$, which is at most 1 since the resolvent of a monotone operator is nonexpansive \cite{boyd}. Then
\begin{align}  \label{eq:psi-lip}
Lip(\psi) \leq 2 C_1 k d_1 \|R-I\| .
\end{align}
Hence there exists $\delta_1 > 0$ such that if $\|R-I\| < \delta_1$, then $Lip(\psi) < 1$, in which case Lemma \ref{eq:liplemma} implies $\Phi_R = (I+\psi)^{-1}$ is well-defined with $Lip(\Phi_R) \leq 1/(1-Lip(\psi))$.  Lemma~\ref{eq:liplemma} with $\Psi = I + \psi = \Phi_R^{-1}$ implies
\begin{align}
    Lip(I-\Phi_R) =Lip(I-\Psi^{-1}) \leq \frac{Lip(\psi)} {1-Lip(\psi)}.
\end{align}
By \eqref{eq:psi-lip}, we can choose $\delta > 0$ so that if $\|R-I\| < \delta$, then $Lip(I-\Phi_R) \leq C \|R-I\|$, where $C = 4 C_1 k d_1$. 

Recall that $\phi$ strongly monotone means that there exists $m > 0$ so that for all $x,v$, $(x-v)^T(\phi(x)-\phi(v)) \geq m ||x-v||^2_2$.  Let $\eta = \tilde{R}^{-1} - I$.  By \eqref{eq:I-tildeR}, we can reduce $\delta$ to get $\|\eta\| < \frac{m} {2k}$. To show that $\tilde{R}^{-1}\phi$ is maximal monotone, note that
\begin{align*}
    (x&-v)^T (\tilde{R}^{-1}\phi(x) - \tilde{R}^{-1}\phi(v)) \nonumber\\
    &= (x-v)^T(\phi(x) - \phi(v)) - (x-v)^T(\eta(\phi(x)-\phi(v)) 
\end{align*}
By assumption, the first term is bounded below by $m ||x-v||^2_2$. Additionally $\|\phi(x)-\phi(v)\| \leq  k\|x-v\|$ by assumption, so
\begin{align}
    (x-v)^T&(\tilde{R}^{-1}\phi(x) - \tilde{R}^{-1}\phi(v)) \\
    &\geq m ||x-v||^2_2 - k \|\eta\| \|x-v\|_2^2,
\end{align}
which gives a lower bound of $\frac{m} {2} \|x-v\|_2^2$. Hence $\tilde{R}^{-1}\phi$ is strongly monotone, and since  $\tilde{R}^{-1}$ is linear, $\tilde{R}^{-1}\phi$ is maximal monotone.
\end{proof}

\vspace{0.1in}
\begin{thm}  \label{thm:RAP-equiv} 
Suppose $\phi$ is a Lipschitz and strongly maximal monotone function and let $H = (I + \phi)^{-1}$ and $\mu_1 = \mu_2 = 1/2$. Assume $\sigma^2 < 1/L^2$.  
Then there exist $\alpha > 0$ and $C > 0$ such that if $R$ is a matrix satisfying $R A^T = B$ and $\|R-I\| \leq \alpha < 1$, there exists $\Phi_R$ a Lipschitz map depending on $H$ and $R$ with $\text{Lip}(\Phi_R - I) \leq C \|R-I\|$ such that the following two choices lead to the same set of solutions $x^*$ in \eqref{eq:FG}:
\begin{itemize}
    \item $\textbf{F}_1 = \tilde{F}$ is the RAP update in \eqref{eq:mismatch} and $\textbf{F}_2 = H$;
    \item $\textbf{F}_1 = F$ is the standard update in \eqref{eq:5} and $\textbf{F}_2 = H \circ \Phi_R$.  
\end{itemize} 
\end{thm}

\medskip
This theorem says that the effect of RAP with a denoiser $H$ can be explained by using the standard data-fitting term together with a modified denoiser defined by a Lipschitz transformation of the image domain followed by the original denoiser.  

\medskip
\begin{proof}
In order to apply Theorem 1, we verify that each $\textbf{F}_j$ is of the form $(I+\omega)^{-1}$ for $\omega$ a maximal monotone function.
By Lemma~\ref{eq:lemmaFform}, we have that $F = (I+\nabla f)^{-1}$ and $\nabla f$ is maximal monotone. As $\phi$ is maximal monotone, $H$ is of the desired form, and $H$ is well-defined by Proposition~\ref{propn:boyd}. 
Since $\sigma^2 < 1/L^2$, Lemma~\ref{eq:lemmaFRform} with the same $f,r$ as defined in Lemma~\ref{eq:lemmaFform} implies that $\tilde{F} = (I+\tilde{R}\nabla f)^{-1}$. Lemma~\ref{eq:lemmaFRform} also gives that $\tilde{R}\nabla f$ is maximal monotone. 
Finally as $\phi$ is assumed to be a Lipschitz and strongly maximal function, Lemma~\ref{eq:lemmaphiRform} gives that $H \circ \Phi_R = (I+\tilde{R}^{-1}\phi)^{-1}$ is well-defined and that $\tilde{R}^{-1}\phi$ is maximal monotone.
Thus we may apply the results of Theorem 1 to each pair $(\tilde F, H)$ and $(F, H \circ \Phi_R)$. 

From the proof of Theorem 1, since $\tilde F = (I + \tilde R \nabla f)^{-1}$ and $H = (I + \phi)^{-1}$ we see that $\mathbf{v^*} = (v_1^*, v_2^*)$ is a solution of 
$$ \begin{bmatrix} \tilde F(v_1^*)\\ H(v_2^*)
\end{bmatrix} = \mathbf{G}(\mathbf{v^*})
$$
if and only if 
$$ \tilde R \nabla f(x^*) + \phi(x^*) = 0, $$
where $x^*$ is the common entry of the stacked vector $\mathbf{G}(\mathbf{v^*})$.
Since $\tilde R$ is invertible, this is equivalent to
$$ \nabla f(x^*) + \tilde R^{-1} \phi(x^*) = 0. $$
Since $F = (I + \nabla f)^{-1}$ and $ H \circ \Phi_R = (I + \tilde R^{-1} \phi)^{-1}$, again the proof of Theorem 1 implies that this is true if and only if $\mathbf{\tilde v^*} = (\tilde v_1^*, \tilde v_2^*)$ is a solution of
$$ \begin{bmatrix} F(\tilde v_1^*)\\ H(\Phi_R(\tilde v_2^*))
\end{bmatrix} = \mathbf{G}(\mathbf{\tilde v^*})
$$
with $x^*$ the common entry of the stacked vector $\mathbf{G}(\mathbf{\tilde v^*})$.

This implies that the two formulations have the same set of consensus solutions, $x^*$.  

\end{proof}

\begin{propn}
If $A$ is an $n \times n$ symmetric matrix with eigenvalues in $(0,1]$ and $b \in {\mathbb R}^n$, then the mapping $F(x) = Ax + b$ is a proximal map. 
\end{propn}

\medskip
\begin{proof}
The conditions on $A$ imply that $A^{-1} - I$ is symmetric and positive semidefinite, so the Cholesky decomposition gives an invertible $R$ such that $R^TR = \frac{1} {\sigma^2}(A^{-1}-I)$ (for specified $\sigma^2 > 0$).  
Define $p$ so that $\sigma^2 A R^T p = b$ and consider the proximal map defined by
\begin{align}
    \argmin_u \left \{ \frac{1} {2} ||Ru-p||^2 + \frac{1} {2\sigma^2} ||u-x||^2 \right\}.  
\end{align}
The first-order optimality condition yields
\begin{align}
    R^T(Ru^*-p) + \frac{1} {\sigma^2}(u^*-x) = 0,
\end{align}
and gathering the $u^*$ terms and multiplying by $\sigma^2$ gives
\begin{align}
    (I + \sigma^2 R^T R) u^* = x + \sigma^2 R^T p.
\end{align}
Noting that $(I+ \sigma^2 R^T R) = A^{-1}$ and using the choice of $p$ gives $u^* = Ax + b$.  Hence $F(x) = Ax+b$ is a proximal map.
\end{proof}

\vspace{0.1in}

\begin{propn} \label{prop:W-prox}
Let $F(x) = Wx+q$ where $W = V \Lambda V^{-1}$ with $\Lambda$ diagonal having eigenvalues in $(0,1]$ and $q \in \mathbb{R}^N$. For $x \in \mathbb{R}^N$ define $\hat{x} = V^{-1}x$. Then $\hat F ( \hat x ) = V^{-1} F(V \hat x)$ is a proximal map in the coordinates $\hat{x}$.
\end{propn}

\medskip
\begin{proof}
Expanding $\hat F$ using $W = V \Lambda V^{-1}$ gives $\hat F ( \hat x) = \Lambda \hat x + V^{-1} q $. Since $\Lambda$ is diagonal with eigenvalues in $(0,1]$, the previous proposition implies that $\hat F( \hat x)$ is a proximal map. 
\end{proof}

\vspace{0.1in}

Proposition~3 applies with $F$ as the RAP update in \eqref{eq:mismatch} and $\hat F$ as the standard update in \eqref{eq:5}.  Theorem~3 then implies that Algorithm 1 converges to a fixed point using the RAP update.

\begin{thm} \label{thm:RAPconv}
Let $F$ and $\hat F$ be as in Proposition~\ref{prop:W-prox}.  Let $H$ be a denoiser such that $\hat H(\hat x) = V^{-1}H(V \hat x)$ is nonexpansive in the coordinates $\hat{x}$. Then Algorithm 1 converges using the operators $F$ and $H$. 
\end{thm}

\medskip

\begin{proof}
An expansion of $\mathbf{F}$ and $\mathbf{G}$ shows that Algorithm~1 with two operators is equivalent to the standard PnP algorithm of \cite{venkatakrishnan_plug-and-play_2013}.
By Proposition~\ref{prop:W-prox}, $\hat F$ is a proximal map, and by assumption, $\hat H$ is nonexpansive. Hence by \cite{buzzard_plug-and-play_2018}, Algorithm ~1 using operators $\mathbf{F}_1 = \hat F$ and $\mathbf{F}_2 = \hat H$ converges to a fixed point.  

The bilinear change of variables $\hat x = V^{-1}x$ yields a one-to-one correspondence 
\begin{align}
    \hat F (\hat x) = V^{-1} F( V \hat x) \\
    \hat H (\hat x) = V^{-1} H( V \hat x) 
\end{align}
Applying this to each component and map in Algorithm~1 produces a shadow sequence equivalent to running Algorithm~1 with $\mathbf{F}_1 = F$ and $\mathbf{F}_2 = H$.  This shadow sequence converges by continuity of $V$ and $V^{-1}$, so the PnP algorithm converges using $F$ and $H$.
\end{proof}


%

\section*{Acknowledgment}

This work was supported by UES Inc. under prime contract FA8650-15- D-5405 and by NSF grant CCF-1763896. The authors thank Cheri M. Hampton for assistance in data acquisition and Asif Mehmood for his mentorship.

\ifCLASSOPTIONcaptionsoff
  \newpage
\fi



%
\printbibliography

@inproceedings{anderson_sparse_2013,
	title = {Sparse imaging for fast electron microscopy},
	volume = {8657},
	doi = {10.1117/12.2008313},
	booktitle = {Computational {Imaging} {XI}},
	publisher = {International Society for Optics and Photonics},
	author = {Anderson, Hyrum S. and Ilic-Helms, Jovana and Rohrer, Brandon and Wheeler, Jason and Larson, Kurt},
	month = feb,
	year = {2013},
	pages = {86570C}
}

@misc{frc_code,
 author = { Ali, Sajid},
 year   = {2018},
 title  = {Fourier Ring Correlation},
 note   = {\url{https://github.com/s-sajid-ali/FRC}}
}

@article{park_nanorods,
    title={Highly Concentrated Seed-Mediated Synthesis of Monodispersed Gold Nanorods},
    author={Kyoungweon Park and Ming-siao Hsiao and Yoon-Jae Yi and Sarah Izor and Hilmar Koerner and Ali Jawaid and Richard A. Vaia}, 
    journal={ACS Applied Materials \& Interfaces},
    year={2017}, 
    volume={9},
    number={31},
    pages={26363-26371},
    doi = {10.1021/acsami.7b08003}
}

@article{he_construction_2021,
	title = {Construction of 3-{D} realistic representative volume element failure prediction model of high density rigid polyurethane foam treated under complex thermal-vibration conditions},
	volume = {193},
	doi = {10.1016/j.ijmecsci.2020.106164},
	journal = {International Journal of Mechanical Sciences},
	author = {He, Yannan and Wu, Jiacheng and Qiu, Dacheng and Yu, Zhiqiang},
	month = mar,
	year = {2021},
	pages = {106164},
}

@article{bouman_plug-and-play_2021,
	title = {Plug-and-{Play}: {A} {General} {Approach} for the {Fusion} of {Sensor} and {Machine} {Learning} {Models}},
	volume = {54},
	shorttitle = {Plug-and-{Play}},
	number = {2},
	journal = {SIAM News},
	author = {Bouman, Charles A. and Buzzard, Gregery T. and Wohlberg, Brendt},
	month = mar,
	year = {2021},
	pages = {1,4},
}

@article{antun_instabilities_2020,
	title = {On instabilities of deep learning in image reconstruction and the potential costs of {AI}},
	volume = {117},
	copyright = {© 2020 . https://www.pnas.org/site/aboutpnas/licenses.xhtmlPublished under the PNAS license.},
	doi = {10.1073/pnas.1907377117},
	number = {48},
	journal = {Proceedings of the National Academy of Sciences},
	author = {Antun, Vegard and Renna, Francesco and Poon, Clarice and Adcock, Ben and Hansen, Anders C.},
	month = dec,
	year = {2020},
	pmid = {32393633},
	note = {Publisher: National Academy of Sciences
Section: Colloquium on the Science of Deep Learning},
	pages = {30088--30095},
}

@inproceedings{shocher_zero-shot_2018,
	address = {Salt Lake City, UT},
	title = {Zero-{Shot} {Super}-{Resolution} {Using} {Deep} {Internal} {Learning}},
	doi = {10.1109/CVPR.2018.00329},
	booktitle = {2018 {IEEE}/{CVF} {Conference} on {Computer} {Vision} and {Pattern} {Recognition}},
	publisher = {IEEE},
	author = {Shocher, Assaf and Cohen, Nadav and Irani, Michal},
	month = jun,
	year = {2018},
	pages = {3118--3126}
}

@inproceedings{venkatakrishnan_plug-and-play_2013,
	title = {Plug-and-{Play} priors for model based reconstruction},
	doi = {10.1109/GlobalSIP.2013.6737048},
	booktitle = {2013 {IEEE} {Global} {Conference} on {Signal} and {Information} {Processing}},
	author = {Venkatakrishnan, Singanallur V. and Bouman, Charles A. and Wohlberg, Brendt},
	month = dec,
	year = {2013},
	pages = {945--948}
}

@inproceedings{dong_learning_2014,
	address = {Cham},
	series = {Lecture {Notes} in {Computer} {Science}},
	title = {Learning a {Deep} {Convolutional} {Network} for {Image} {Super}-{Resolution}},
	doi = {10.1007/978-3-319-10593-2_13},
	booktitle = {Computer {Vision} – {ECCV} 2014},
	publisher = {Springer International Publishing},
	author = {Dong, Chao and Loy, Chen Change and He, Kaiming and Tang, Xiaoou},
	editor = {Fleet, David and Pajdla, Tomas and Schiele, Bernt and Tuytelaars, Tinne},
	year = {2014},
	pages = {184--199}
}

@inproceedings{ledig_photo-realistic_2017,
	title = {Photo-{Realistic} {Single} {Image} {Super}-{Resolution} {Using} a {Generative} {Adversarial} {Network}},
	doi = {10.1109/CVPR.2017.19},
	booktitle = {2017 {IEEE} {Conference} on {Computer} {Vision} and {Pattern} {Recognition} ({CVPR})},
	author = {Ledig, Christian and Theis, Lucas and Huszár, Ferenc and Caballero, Jose and Cunningham, Andrew and Acosta, Alejandro and Aitken, Andrew and Tejani, Alykhan and Totz, Johannes and Wang, Zehan and Shi, Wenzhe},
	month = jul,
	year = {2017},
	pages = {105--114}
}

@inproceedings{sajjadi_enhancenet_2017,
	title = {{EnhanceNet}: {Single} {Image} {Super}-{Resolution} {Through} {Automated} {Texture} {Synthesis}},
	shorttitle = {{EnhanceNet}},
	doi = {10.1109/ICCV.2017.481},
	booktitle = {2017 {IEEE} {International} {Conference} on {Computer} {Vision} ({ICCV})},
	author = {Sajjadi, Mehdi S. M. and Schölkopf, Bernhard and Hirsch, Michael},
	month = oct,
	year = {2017},
	pages = {4501--4510}
}

@inproceedings{sreehari_multi-resolution_2017,
	title = {Multi-{Resolution} {Data} {Fusion} for {Super}-{Resolution} {Electron} {Microscopy}},
	doi = {10.1109/CVPRW.2017.146},
	booktitle = {2017 {IEEE} {Conference} on {Computer} {Vision} and {Pattern} {Recognition} {Workshops} ({CVPRW})},
	author = {Sreehari, Suhas and Venkatakrishnan, S. V. and Bouman, Katherine L. and Simmons, Jeffrey P. and Drummy, Lawrence F. and Bouman, Charles A.},
	month = jul,
	year = {2017},
	pages = {1084--1092}
}

@inproceedings{yamanaka_fast_2018,
author = {Yamanaka, Jin and Kuwashima, Shigesumi and Kurita, Takio},
year = {2017},
month = {10},
pages = {217-225},
title = {Fast and Accurate Image Super Resolution by Deep CNN with Skip Connection and Network in Network},
doi = {10.1007/978-3-319-70096-0_23}
}

@article{zhang_beyond_2017,
  author={Zhang, Kai and Zuo, Wangmeng and Chen, Yunjin and Meng, Deyu and Zhang, Lei},
  journal={IEEE Transactions on Image Processing}, 
  title={Beyond a Gaussian Denoiser: Residual Learning of Deep CNN for Image Denoising}, 
  year={2017},
  volume={26},
  number={7},
  pages={3142-3155},
  doi={10.1109/TIP.2017.2662206}}

@article{buzzard_plug-and-play_2018,
	title = {Plug-and-{Play} {Unplugged}: {Optimization}-{Free} {Reconstruction} {Using} {Consensus} {Equilibrium}},
	volume = {11},
	shorttitle = {Plug-and-{Play} {Unplugged}},
	doi = {10.1137/17M1122451},
	number = {3},
	journal = {SIAM Journal on Imaging Sciences},
	author = {Buzzard, Gregery T. and Chan, Stanley H. and Sreehari, Suhas and Bouman, Charles A.},
	month = jan,
	year = {2018},
	note = {Publisher: Society for Industrial and Applied Mathematics}
}

@inproceedings{rakotonirina_esrgan_2020,
	title = {{ESRGAN}+ : {Further} {Improving} {Enhanced} {Super}-{Resolution} {Generative} {Adversarial} {Network}},
	shorttitle = {{ESRGAN}+},
	doi = {10.1109/ICASSP40776.2020.9054071},
	booktitle = {{ICASSP} 2020 - 2020 {IEEE} {International} {Conference} on {Acoustics}, {Speech} and {Signal} {Processing} ({ICASSP})},
	author = {Rakotonirina, Nathanaël Carraz and Rasoanaivo, Andry},
	month = may,
	year = {2020},
	pages = {3637--3641}
}

@inproceedings{majee_4d_2019,
	title = {{4D} {X}-{Ray} {CT} {Reconstruction} using {Multi}-{Slice} {Fusion}},
	doi = {10.1109/ICCPHOT.2019.8747328},
	booktitle = {2019 {IEEE} {International} {Conference} on {Computational} {Photography} ({ICCP})},
	author = {Majee, Soumendu and Balke, Thilo and Kemp, Craig A. J. and Buzzard, Gregery T. and Bouman, Charles A.},
	month = may,
	year = {2019},
	pages = {1--8}
}

@inproceedings{zhang_deep_2019,
	title = {Deep {Plug}-{And}-{Play} {Super}-{Resolution} for {Arbitrary} {Blur} {Kernels}},
	doi = {10.1109/CVPR.2019.00177},
	booktitle = {2019 {IEEE}/{CVF} {Conference} on {Computer} {Vision} and {Pattern} {Recognition} ({CVPR})},
	author = {Zhang, Kai and Zuo, Wangmeng and Zhang, Lei},
	month = jun,
	year = {2019},
	pages = {1671--1681}
}

@InProceedings{wang2018esrgan,
    author = {Wang, Xintao and Yu, Ke and Wu, Shixiang and Gu, Jinjin and Liu, Yihao and Dong, Chao and Qiao, Yu and Loy, Chen Change},
    title = {ESRGAN: Enhanced super-resolution generative adversarial networks},
    booktitle = {The European Conference on Computer Vision Workshops (ECCVW)},
    month = {9},
    year = {2018}
}

@article{lalush_improving_1994,
	title = {Improving the convergence of iterative filtered backprojection algorithms},
	volume = {21},
	copyright = {© 1994 American Association of Physicists in Medicine},
	doi = {10.1118/1.597210},
	number = {8},
	journal = {Medical Physics},
	author = {Lalush, David S. and Tsui, Benjamin M. W.},
	year = {1994},
	pages = {1283--1286},
}

@inproceedings{ziabari_mismatch_2018,
author = {Ziabari, Amirkoushyar and Ye, Dong Hye and Fu, Lin and Srivastava, Somesh and Sauer, Ken and Thibault, Jean-Baptist and Bouman, Charles},
year = {2018},
booktitle = {The 5th International Conference on Image Formation in X-Ray Computed Tomography},
month = {12},
pages = {15-18},
title = {Model Based Iterative Reconstruction With Spatially Adaptive Sinogram Weights for Wide-Cone Cardiac CT}
}

@INPROCEEDINGS{7532589,  author={A. {Brifman} and Y. {Romano} and M. {Elad}},  booktitle={2016 IEEE International Conference on Image Processing (ICIP)},   title={Turning a denoiser into a super-resolver using plug and play priors},   year={2016},  volume={},  number={},  pages={1404-1408},}

@article{zhang2020plug,
author = {Zhang, Kai and Li, Yawei and Zuo, Wangmeng and Zhang, Lei and Timofte, Radu},
title = {Plug-and-Play Image Restoration with Deep Denoiser Prior},
journal = {arXiv:2008.13751},
year = {2020},
}

@inproceedings{zhang2017learning,
   title={Learning Deep CNN Denoiser Prior for Image Restoration},
   author={Zhang, Kai and Zuo, Wangmeng and Gu, Shuhang and Zhang, Lei},
   booktitle={IEEE Conference on Computer Vision and Pattern Recognition},
   pages={3929--3938},
   year={2017},
 }

@article{SHI2020,
title = "Deep prior-based sparse representation model for diffraction imaging: A plug-and-play method",
journal = "Signal Processing",
volume = "168",
pages = "107350",
year = "2020",
author = "Baoshun Shi and Qiusheng Lian and Huibin Chang",
}

@article{schofield_image_2020,
	title = {Image reconstruction: {Part} 1 – understanding filtered back projection, noise and image acquisition},
	volume = {14},
	doi = {https://doi.org/10.1016/j.jcct.2019.04.008},
	number = {3},
	journal = {Journal of Cardiovascular Computed Tomography},
	author = {Schofield, R. and King, L. and Tayal, U. and Castellano, I. and Stirrup, J. and Pontana, F. and Earls, J. and Nicol, E.},
	year = {2020},
	pages = {219 -- 225}
}

@ARTICLE{ghani_data_2020,
  author={Ghani, Muhammad Usman and Karl, W. Clem},
  journal={IEEE Transactions on Computational Imaging}, 
  title={Data and Image Prior Integration for Image Reconstruction Using Consensus Equilibrium}, 
  year={2021},
  volume={7},
  number={},
  pages={297-308},
  doi={10.1109/TCI.2021.3062986}}

@techreport{chouzenoux:hal-02961431,
  TITLE = {{Convergence of Proximal Gradient Algorithm in the Presence of Adjoint Mismatch}},
  AUTHOR = {Chouzenoux, Emilie and Pesquet, Jean-Christophe and Riddell, Cyril and Savanier, Marion and Trousset, Yves},
  URL = {https://hal.archives-ouvertes.fr/hal-02961431},
  TYPE = {Research Report},
  INSTITUTION = {{CentraleSup{\'e}lec}},
  YEAR = {2020},
  MONTH = Oct,
  HAL_ID = {hal-02961431},
  HAL_VERSION = {v1},
}

@ARTICLE{splines,
  author={S. A. {Dyer} and J. S. {Dyer}},
  journal={IEEE Instrumentation   Measurement Magazine}, 
  title={Cubic-spline interpolation. 1}, 
  year={2001},
  volume={4},
  number={1},
  pages={44-46},
  doi={10.1109/5289.911175}}

@article{nlmissues,
author = {Liu, Baozhong and Liu, Jianbin},
year = {2018},
month = {01},
pages = {03029},
title = {Overview of image noise reduction based on non-local mean algorithm},
volume = {232},
journal = {MATEC Web of Conferences},
doi = {10.1051/matecconf/201823203029}
}

@article{Qian_2020,
   title={Effective Super-Resolution Methods for Paired Electron Microscopic Images},
   volume={29},
   DOI={10.1109/tip.2020.3000964},
   journal={IEEE Transactions on Image Processing},
   publisher={Institute of Electrical and Electronics Engineers (IEEE)},
   author={Qian, Yanjun and Xu, Jiaxi and Drummy, Lawrence F. and Ding, Yu},
   year={2020},
   pages={7317–7330}
}

@ARTICLE{nairconv,
  author={P. {Nair} and R. G. {Gavaskar} and K. N. {Chaudhury}},
  journal={IEEE Transactions on Computational Imaging}, 
  title={Fixed-Point and Objective Convergence of Plug-and-Play Algorithms}, 
  year={2021},
  volume={7},
  number={},
  pages={337-348},
  doi={10.1109/TCI.2021.3066053}}

@article{boyd,
author = {Ryu, E. K. and Boyd, S.},
title = {Primer on Monotone Operator Methods},
journal = {Applied and Computational Mathematics},
volume = {15},
number = {1},
pages = {3 -- 43}, 
year = {2016},
}

@inproceedings{bree-SEM,
author = {Bree, P.J. and van Lierop, C. M. M. and Bosch, P.P.J.},
year = {2010},
month = {08},
pages = {268 - 273},
title = {On hysteresis in magnetic lenses of electron microscopes},
journal = {IEEE International Symposium on Industrial Electronics},
doi = {10.1109/ISIE.2010.5637564}
}

@article{clarke,
author = {F. H. Clarke},
title = {{On the inverse function theorem.}},
volume = {64},
journal = {Pacific Journal of Mathematics},
number = {1},
publisher = {Pacific Journal of Mathematics, A Non-profit Corporation},
pages = {97 -- 102},
year = {1976},
doi = {pjm/1102867214},
}

@article{multimodalMR,
author = {Marti-Bonmati, Luis and Sopena, Ramón and Bartumeus, Paula and Sopena, Pablo},
year = {2010},
month = {07},
pages = {180-9},
title = {Multimodality imaging techniques},
volume = {5},
journal = {Contrast media \& molecular imaging},
doi = {10.1002/cmmi.393}
}

@ARTICLE{gaoSR,
  author={Gao, Lianru and Hong, Danfeng and Yao, Jing and Zhang, Bing and Gamba, Paolo and Chanussot, Jocelyn},
  journal={IEEE Transactions on Geoscience and Remote Sensing}, 
  title={Spectral Superresolution of Multispectral Imagery With Joint Sparse and Low-Rank Learning}, 
  year={2021},
  volume={59},
  number={3},
  pages={2269-2280},
  doi={10.1109/TGRS.2020.3000684}}

@ARTICLE{hongSAR,
  author={Hong, Danfeng and Gao, Lianru and Yokoya, Naoto and Yao, Jing and Chanussot, Jocelyn and Du, Qian and Zhang, Bing},
  journal={IEEE Transactions on Geoscience and Remote Sensing}, 
  title={More Diverse Means Better: Multimodal Deep Learning Meets Remote-Sensing Imagery Classification}, 
  year={2021},
  volume={59},
  number={5},
  pages={4340-4354},
  doi={10.1109/TGRS.2020.3016820}}

@article{ALAM201924,
title = {Challenges and Solutions in Multimodal Medical Image Subregion Detection and Registration},
journal = {Journal of Medical Imaging and Radiation Sciences},
volume = {50},
number = {1},
pages = {24-30},
year = {2019},
issn = {1939-8654},
doi = {https://doi.org/10.1016/j.jmir.2018.06.001},
author = {Fakhre Alam and Sami Ur Rahman},
}

@article{moselyAHA,
author = {Michael Moseley  and Geoffrey Donnan },
title = {Multimodality Imaging},
journal = {Stroke},
volume = {35},
number = {11\_suppl\_1},
pages = {2632-2634},
year = {2004},
doi = {10.1161/01.STR.0000143214.22567.cb},
}

@INPROCEEDINGS{venkatBP,
  author={Venkatakrishnan, S. V. and Cakmak, Ercan and Billheux, Hassina and Bingham, Philip and Archibald, Richard K.},
  booktitle={2017 51st Asilomar Conference on Signals, Systems, and Computers}, 
  title={Model-based iterative reconstruction for neutron laminography}, 
  year={2017},
  volume={},
  number={},
  pages={1864-1869},
  doi={10.1109/ACSSC.2017.8335686}}

@article{BanterleFRC,
title = {Fourier ring correlation as a resolution criterion for super-resolution microscopy},
journal = {Journal of Structural Biology},
volume = {183},
number = {3},
pages = {363-367},
year = {2013},
issn = {1047-8477},
doi = {https://doi.org/10.1016/j.jsb.2013.05.004},
author = {Niccolò Banterle and Khanh Huy Bui and Edward A. Lemke and Martin Beck},
}

@article{sreehari2016plug,
  title={Plug-and-play priors for bright field electron tomography and sparse interpolation},
  author={S. Sreehari and S. V. Venkatakrishnan and B. Wohlberg and G. T. Buzzard and L. F. Drummy and J. P. Simmons and C. A. Bouman},
  journal={Transactions on Computational Imaging},
  volume={2},
  pages={408-423},
  month= {},
  year={2016},
}

@InProceedings{yao2020cross_attention,
author="Yao, Jing
and Hong, Danfeng
and Chanussot, Jocelyn
and Meng, Deyu
and Zhu, Xiaoxiang
and Xu, Zongben",
editor="Vedaldi, Andrea
and Bischof, Horst
and Brox, Thomas
and Frahm, Jan-Michael",
title="Cross-Attention in Coupled Unmixing Nets for Unsupervised Hyperspectral Super-Resolution",
booktitle="Computer Vision -- ECCV 2020",
year="2020",
publisher="Springer International Publishing",
address="Cham",
pages="208--224",
isbn="978-3-030-58526-6"
}

%








\end{document}